\documentclass[12pt,preprint]{aastex}
\usepackage{amsmath}
\shorttitle{Gravitational Darkening in Be Stars}
\shortauthors{McGill et al.}

\newcommand{\msun}{\mbox{$M_{\odot}$}}

\newcommand{\rsun}{\mbox{$R_{\odot}$}}

\newcommand{\lsun}{\mbox{$L_{\odot}$}}

\newcommand{\degang}{\mbox{$^{\circ}$}}

\begin{document} 

\title{The Thermal Structure of Gravitationally-Darkened Classical Be Star Disks}

\author{M.\ A.\ M{c}Gill\altaffilmark{1}, T.\ A.\ A.\ Sigut\altaffilmark{1}, \& C.\ E.\  
Jones\altaffilmark{1}, }

\altaffiltext{1}{Department of Physics and Astronomy, The University 
of Western Ontario, London, Ontario, N6A 3K7, Canada}

\slugcomment{Accepted for Publication in ApJ August 29, 2011}

%
%
\begin{abstract}

The effect of gravitational darkening on models of the thermal structure
of Be star disks is systematically studied for a wide range of Be star
spectral types and rotation rates.  Gravitational darkening causes a
reduction of the stellar effective temperature towards the equator and
a redirection of energy towards the poles. It is an important physical
effect in these star-disk systems because the photoionizing radiation from
the central B~star is the main energy source for the disk.  We have added
gravitational darkening to the {\sc bedisk} code to produce circumstellar
disk models that include both the variation in the effective temperature
with latitude and the non-spherical shape of the star in the calculation
of the stellar photoionizing radiation field.  The effect of gravitational
darkening on global measures of disk temperature is generally significant
for rotation rates above 80\% of critical rotation. For example, a B0V
model rotating at 95\% of critical has a density-averaged disk temperature
$\approx\!2500$~K cooler than its non-rotating counterpart. However,
detailed differences in the temperature structure of disks surrounding
rotating and non-rotating stars reveal a complex pattern of heating and
cooling.  Spherical gravitational darkening, an approximation that ignores
the changing shape of the star, gives good results for disk temperatures
for rotation rates less than $\approx80$\% of critical. However for the
highest rotation rates, the distortion of the stellar surface caused by
rotation becomes important.

\end{abstract}

\keywords{stars: circumstellar matter 
-- stars: emission line, Be \\ 
-- stars: rotation}

\section{Introduction}
\label{intro}

A classical Be-star is a non-supergiant B star that possesses a gaseous
equatorial disk \citep{por03}.  This circumstellar material produces
an emission line spectrum including a prominent ${\rm H} {\alpha}$
emission line.  Some Be stars, for example $\gamma$ Cassiopeia, produce
a rich emission line spectrum with many hydrogen lines and some metal
lines.  Be star disks also produce a continuum excess due to bound-free
and free-free emission.  This excess has been observed at a variety
of wavelengths from the visible to the radio \citep{cot87,wat87}.
Linear polarization of up to $2$\% has been observed in the continuum
emission and is caused by electron scattering in the non-spherical
circumstellar envelope \citep{mb78,poe79}.  Be star disks have been
partially resolved by long-baseline interferometric measurements,
initially in the radio \citep{dou92}, and later at infrared and optical
wavelengths \citep[]{ste01,tyc04}.

The stellar absorption lines of Be stars indicate rapid rotation of the
central star, and this property is thought to play an important role in
the release of material into the disk \citep{por96}. There is strong
evidence that Be star disks are in Keplerian rotation \citep{hum00,oud08};
however, it is still uncertain if the equatorial rotation speeds of Be
stars are close to ``critical", i.e.\ close to the Keplerian orbital
speed at the inner edge of the disk \citep{cra05}.  In addition,
another mechanism must somehow allow for the wide range of behaviours
observed in Be stars: some systems have exhibited essentially stable
H$\alpha$ emission for as long as they have been observed, while others
are highly variable with timescales ranging from less than a day to
decades \citep{por03}.

A rapidly rotating star is changed in two ways: the stellar surface is
distorted with the equatorial radius becoming larger than the polar
radius, and the stellar surface temperature acquires a dependence
on stellar latitude with cooler gas at the equator compared to the
hotter pole. These phenomena together are commonly called gravitational
darkening.  The rotational distortion and surface intensity variation
with latitude have direct observational support from interferometric
observations, including the stars Achernar \citep{des03} and Altair
\citep{mon07}.

Because the equatorial region of the photosphere is both the fastest
rotating and the dimmest, it can be difficult to detect this maximally
Doppler-shifted region in integrated light.  This problem can be
compounded by the obscuration of the equatorial regions by disk
material. As a result, the bounds on Be star rotation rates remain
contentious \citep{cha01, tow04, cra05, fre05}. Nevertheless, the
canonical result is that most Be stars rotate at $\approx\!80$\% of
their critical speed (as defined in Section \ref{theory}). Unfortunately,
it seems that directly measuring projected rotational velocities above
80\% of the critical speed may not be possible due to the dimming of the
equatorial regions \citep[]{tow04}.  \citet{cra05} presents a detailed
statistical analysis of Be star $v\sin i$ values in which he attempts to
match the observed distributions using a parametrized distribution of
equatorial velocities. He concludes that early-type Be stars, O7--B3.5,
must rotate at rates significantly less then their critical speeds
(peaking at 40-60\%) while later type Be stars, B3.5--A0, rotate much
closer to their critical speeds (peaking at 70-90\%).

Previously, the effects of gravitational darkening have been
included in disk models for the stars Achernar, or $\alpha$ Eridani,
\citep{car08_aEri} and $\zeta$ Tauri \citep{car09_zTau}.  However, the
rotating star was assumed to be a spheroid rather than following a Roche
profile (see Section~\ref{theory}).  In the literature, gravitational
darkening has been included only as a fit parameter in models of
individual stars. A comprehensive study on the how the inclusion of
gravitational darkening affects models of Be star disks has not yet
been performed.

In this paper, we systematically examine the effect of gravitational
darkening on the thermal structure of a circumstellar disk.  We restrict
our investigation to changes in the energy reaching the disk and the
resulting changes in the temperature distribution.  Section~\ref{theory}
outlines the basic theory behind gravitational darkening and also
gives computational details.  Section~\ref{cals} describes the stellar
and disk models chosen for this investigation. Results are presented
in Section~\ref{results} where the consequences of rotation on the
energy reaching the circumstellar disk (Section~\ref{results1})
and the corresponding changes in the disk temperature structure
(Section~\ref{results2}) are investigated.  In Section~\ref{SGD/PSD},
we examine the two principle ingredients of gravitational darkening
(temperature variation with latitude and distortion of the stellar
surface) separately in order to gauge their relative importance. We
also evaluate the effectiveness of the simpler spherical gravitational
darkening approximation in which the shape distortion produced by rapid
rotation is ignored.  Section~\ref{conclusions} gives the conclusions,
and in Appendix A, we discuss the subtle aspect of how varying the
equatorial radius can change the inherent properties of our disk models.

\section{Theory of Gravitational Darkening}
\label{theory}

\subsection{The Central Star}

Gravitational darkening, first described by von Zeipel (1924), is due to a rotational, 
centrifugal term included in the classical gravitational potential.  For a star 
rotating at a fixed angular speed $\omega$, the potential, $\Phi$, in spherical 
coordinates ($r$,~$\theta$,~$\phi$), is given by
\begin{equation}
\label{total_pot}
\Phi \left( r, \theta \right) =
 - \frac{GM
}{ r } - \frac{1}{2} \omega^2 r^2 \sin^2\theta,
\end{equation}
\noindent where \begin{math} G \end{math} is the gravitational constant, \begin{math} M
\end{math} is the stellar mass, and \begin{math} \theta \end{math} is the stellar co-latitude 
($\theta = 0^o$ on the polar axis). 
The local gravitational acceleration is given by
\begin{eqnarray}
\label{g}
\vec{g} \left( \theta \right) =\left( {\omega^2}r\sin^2 \! \! \theta-\frac{GM
}{ r^2 }\right) \! \hat{r} 
+ \left( \omega^2r\sin \! \theta\cos\!\theta \right)\!\hat{\theta}\;. 
\end{eqnarray}
The angular speed at which the local value of gravity at the equator becomes zero defines the 
critical angular speed, $\omega_{\rm{crit}}$, and the corresponding critical rotation velocity, 
$v_{\rm{crit}}$.  The angular speeds and the equatorial velocities of stars are often expressed 
as fractions of these critical values, $\omega_{\rm{frac}}$ and $v_{\rm{frac}}$.  The relevant 
equations are
\begin{equation}
\begin{split}
\label{definitions}
\omega_{\rm{crit}}&=\sqrt{\frac{8 G M
}{27 r_p^3}},{\hskip 0.2cm} {\omega_{\rm{frac}}} \equiv \frac{\omega}{\omega_{\rm{crit}}},  \\
     v_{\rm{crit}}&=\sqrt{\frac{2 G M
}{ 3 r_p  }},{\hskip 0.2cm}      {v_{\rm{frac}}} \equiv \frac{v_{\rm eq}}{v_{\rm{crit}}} \;,
\end{split}
\end{equation} 
where $r_p$ is the polar radius of the star. In our calculations we have assumed that the polar radius
remains constant following \citet{col66a}.

By requiring the potential across the surface to be constant, 
the radius as a function of $\theta$ is found to be \citep[see again][]{col66a}
\begin{equation}
\label{R_theta}
r(\theta) = \left( \frac{-3  r_{\rm {p}} }{\omega_{\rm{frac}} \sin \theta  } \right) \cos \left( \frac{ \arccos\left(  \omega_{\rm{frac}} \sin \theta  \right)  + 4 \pi}{3} \right) \,.
\end{equation}
When the equator of the star becomes unbound at ${\omega} = \omega_{\rm{crit}}$, 
$ r(\theta=\frac{\pi}{2})$ takes on its largest value, $\frac{3}{2} r_p$. 
The Roche distortion of a stellar surface has direct interferometric confirmation.
For example, \citet{zhao10} present reconstructed images of $\alpha\;$Cep and $\alpha\;$Oph which clearly
reveal rotational distortion consistent with the von Zeipel theory.
The shape of the distorted surface affects the stellar radiation intercepted by the disk as the 
mid-latitudes are tipped towards the pole and away from the equatorial regions.  In addition,
rapid rotation increases the surface area and projected surface area of the star. 
The physical effects associated with near critical rotation have been recently reviewed by
\citet{mey10}.

The local effective temperature of the stellar atmosphere, $T_{\rm eff}$, and the local surface 
gravity, $|\vec{g}|$, are related by von Zeipel's theorem,
\begin{equation} 
\label{Von}
T_{\rm eff}(\theta) = ( C_{\omega} |\vec{g}(\theta)| )^{1/4},
\end{equation}
where $C_{\omega}$ is von Zeipel's constant. \citet{vin07} find that
the von Zeipel variation of $T_{\rm eff}$ with stellar latitude is
required to simultaneously fit several He\,{\sc i} photospheric absorption
profiles from five rapidly rotating, early-type Be stars. Nevertheless,
the power of $1/4$ appearing in the standard gravity-darkening law
has been observationally questioned by \citet{mon07} who find a better
fit to their interferometric observations of Altair with an exponent
of $0.19\pm0.012$. \citet{vanb10} reviews all current interferometric
observations of rapidly rotating stars and finds that of six observed
stars, four are well fit by the standard $1/4$ exponent and two (including
Altair) are not. Despite these results, we will retain the
standard exponent in all calculations to follow.

\begin{figure}
\epsscale{.80}
\plotone{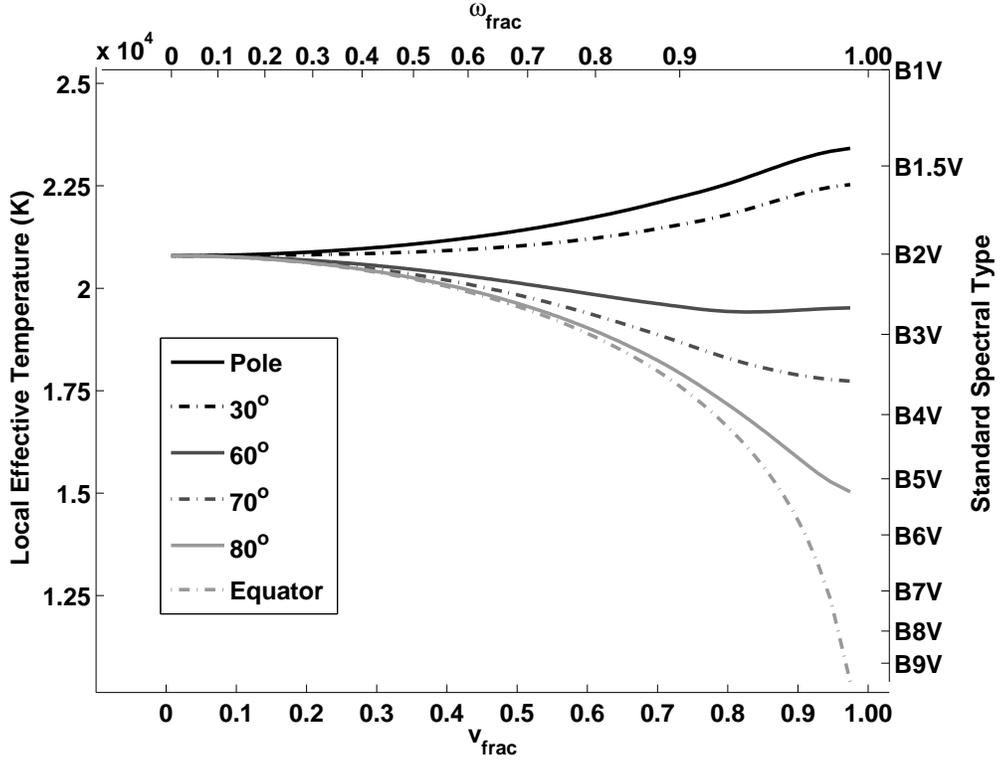}
\caption{Change in the local effective temperature with increasing
$v_{\rm{frac}}$ (lower axis) for a B2V stellar model (see
Table~\ref{star_para}).  The upper axis shows the corresponding
$\omega_{\rm{frac}}$.  The darkest solid line gives the polar temperature
($\theta=0^0$), while the lightest grey dash-dot line shows the equatorial temperature
($\theta=90^0$). Various intermediate stellar latitudes are
also shown.  The right vertical axis shows the standard main sequence
spectral type (with no rotation) for the same surface temperature as
the left axis \citep{cox2000}.  \label{Temperature_profile} } 
\end{figure}

The constant, $C_{\omega}$,  is defined using the assumption that the
total luminosity of the star remains fixed \citep[for discussion of this
point, see][]{lov06}.  Therefore, while the equatorial temperature of a
star decreases with rotation rate, the polar temperature must increase
to maintain a constant luminosity.  This effect is illustrated in
Figure~\ref{Temperature_profile} which shows the change in temperature
with increasing rotation for five stellar co-latitudes:
$0^0$, $30^0$, $60^0$, $70^0$, and $90^0$.  In the absence of rotation,
the stellar surface has a uniform temperature, but as the rotation
speed increases the surface temperatures begin to diverge. Note that
the temperatures of the middle co-latitudes, such as $60^0$, do not
monotonically increase or decrease in temperature. For example as
rotation increases from zero, the temperature at $60^0$ decreases
until $ \sim 0.80\,v_{\rm{crit}}$ where it reaches a minimum
before increasing slightly with increased rotation.

The constant $C_{\omega}$ can be found for a specific set of stellar parameters 
(${r}_p$, ${M}$ and ${L}$), given a value of $\omega$, 
by integrating the local gravity over the stellar surface and 
then multiplying by the luminosity divided by the 
Stephan-Boltzmann constant.  Once $C_{\omega}$ has 
been determined for one set of stellar parameters, the results can be rescaled for any other star 
following \cite{col66a} using
\begin{equation} 
\label{C_scale}
{C^{'}_{\omega}}{C_{\omega}} = \frac { M L^{'} } {M^{'} L} \;.
\end{equation}  
Here $M^{'}$, $L^{'}$, \& $C^{'}_{\omega}$ are the mass, luminosity and
von Zeipel's constant of a particular star, and $M$, $L$, \& $C_{\omega}$ are
for another star with an identical rotation rate.

The local gravity and surface orientation can be established from
Equation~\ref{g}, and the temperature from Equation~\ref{Von}.
This allows the properties of the local stellar atmosphere, $g$ and
$T_{\rm{eff}}$, to be defined for each point on the surface of the star.

\subsection{The Circumstellar Disk}

We now turn to the Be star circumstellar disk.  From a particular
vantage point within the disk, labelled $\vec{s}$, defined using the
cylindrical coordinates, ($s_R,s_z,s_\Phi$), only a portion of the
star is visible.  For a section of the stellar surface specified by
$\vec{r}\left(\theta,\phi\right)$ to be visible from the vantage point,
the dot product between the unit vector pointing from the stellar surface
to the vantage point, $ \hat{n}_{\rm{vp}} = \frac{ \vec{s}-
\vec{r}}{| \vec{r} - \vec{s} |} $, and the local surface normal,
$\hat{n}_{\rm{surf}}=\frac{-\vec{g_{\theta}}}{|\vec{g_{\theta}}|}$,
must be greater then zero. The boundary of this region where
$\hat{n}_{\rm{surf}} \cdot \hat{n}_{\rm{vp}}=0$ is important because it
allows the lines of sight between the star and the vantage point to be
efficiently chosen.
The upper and lower edges of the visible region correspond to the maximum and minimum values of
$\theta$ such that $\hat{n}_{\rm{surf}} \cdot \hat{n}_{\rm{vp}} \ge 0$. These are found 
by solving
\begin{eqnarray}
\label{theta_edge}
\frac{8}{27} \omega^2_{\rm{frac}} \frac{r^3_{\theta}}{r^3_p}  s_{R} \sin \theta 
-  r_{\theta}  - s_{R} \sin\theta + s_{z}\cos \theta = 0
\end{eqnarray} 
for it's two solutions, $\theta_1$ and $\theta_2$.  For values of $\theta$ between $\theta_1$ and 
$\theta_2$, the visible portion of the surface lies between the bounds on $\phi$ given by
\begin{equation}
\label{phi_edge}
 \phi_{\rm{edge}}  = s_{\Phi} \pm 
\arccos \left( \frac{ r(\theta) \sin^2 \theta( 1 -  \kappa_{\theta} ) +  
r(\theta) \cos^2 \theta  - s_{z}  \cos \theta}
{( s_{R} \sin \theta  -  \kappa_\theta s_{R} \sin \theta )} \right),
\end{equation}
with
\begin{equation}
\kappa_\theta = \frac{8}{27}{\omega_{\rm{frac}}^2}\left(\frac{r_{\theta}}{r_{\rm{p}}}\right)^3\,.
\end{equation}

\begin{figure}
\epsscale{.60}
\plotone{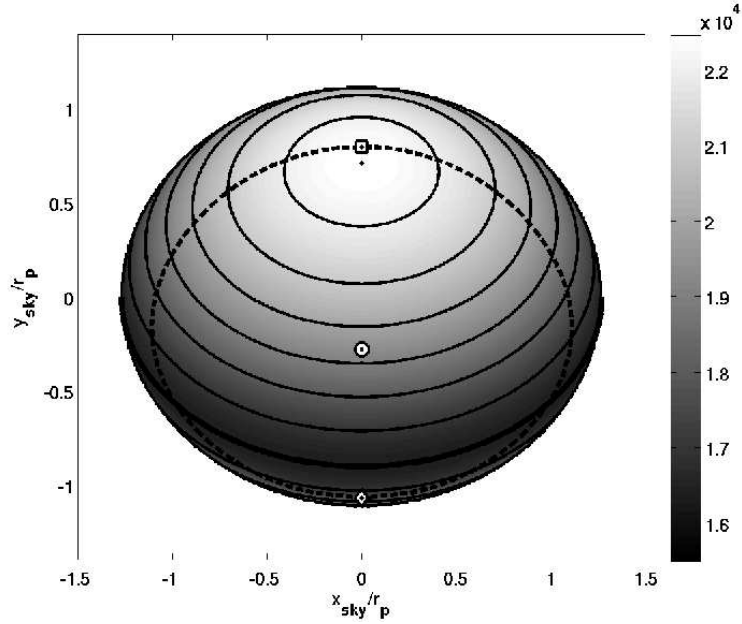}
\caption{The temperature structure of a B2V star rotating at
$v_{\rm{frac}}=0.8$ viewed with an inclination of $45\,^{\circ}$ to the
polar axis (marked by a small dot in uppermost oval).  The bold, solid line is the equator.
The local temperatures shown are determined by Equation~(\ref{Von}).
The dashed line is the limit of the region visible from a vantage point
located within the disk at a radius of $s_{R}=2\,r_p$ and a height of
$s_{z}=1.2\,r_p$, as defined by Equation~(\ref{phi_edge}).  The square
and diamond symbols mark the location of $\theta_1$ and $\theta_2$
respectively which are found by solving Equation~(\ref{theta_edge}).
The direct line of sight from $s_{R}=2\,r_p$ and $s_{z}=1.2\,r_p$ is
indicated by the symbol $\odot$.  \label{star_pictures}}
\end{figure}

Given this geometry, we can now take into account the temperature
changes and the radial distortion of the star to determine the
photoionizing radiation field reaching each point in the disk.
The effects of gravitational darkening are illustrated in
Figure~\ref{star_pictures}.  This figure shows a star rotating at
80\% of $v_{\rm{crit}}$ with its shape distortion and associated surface
temperature variations. Figure~\ref{star_pictures} also shows the visible
region, $\theta_1$, $\theta_2$ and $\phi_{edge}$, for a vantage point at a
radius of $s_{R}=2\,r_p$ and a height above the disk of $s_{z}=1.2\,r_p$.
The disk is assumed to be axisymmetric so any $\Phi$ dependence is not
included.  This figure demonstrates how the visible region changes from
a simple spherical cap for a non-rotating star to an elongated region.

The circumstellar gas is exposed to photoionizing radiation from the
various latitudes of the star and this radiation changes with increasing
rotation.  Different locations within the disk have different regions
of the photosphere within their field of view.  The polar temperature
increases with rotation in contrast to the decrease in the equatorial
temperature (see Figure~\ref{Temperature_profile}).  The drop in the
equatorial temperature occurs at a faster rate than the rise in the
polar temperature.  This means the effects of the hot stellar pole
become significant at speeds above 80\% of $v_{\rm{crit}}$.  In the
equatorial plane up to a distance of $4\,r_p$, stellar latitudes above
25\degang\ are not visible for a circular star. For a critically rotating
star this increases to $\approx$ $30\degang$.  As the gas density in the
disk is assumed to fall off with height above the equatorial plane, the
light reaching upper parts of the disk from the cooler stellar equatorial
region passes through a greater optical depth than that from the hotter
pole.  Thus the effects of gravitational darkening can vary in strength
depending on location within the disk and can have very different effects
on the temperature depending on what part of the disk is considered.
The effect of gravitational darkening cannot be considered to be simply
a reduction of the effective temperature of the star.

\section{Calculations}
\label{cals}

A gravitationally darkened version of {\sc Bedisk} \citep{sig07},
incorporating the theory of Section~\ref{theory}, was created and run for
the stellar parameters given in Table~\ref{star_para} and the rotation
rates given in Table~\ref{v_to_w}.  These stellar parameters
\citep[adopted from][]{cox2000} were chosen to
include a model from four of the five bins used by \citet{cra05}
to analyze the effects of spectral type on Be star rotational
statistics. For reasons discussed below, we have not considered models
from the coolest bin (the bin with characteristic spectral
type B8) considered by \citet{cra05}.
{\sc Bedisk} solves the statistical equilibrium equations
for the  atomic level population equations and then enforces radiative
equilibrium at each point of the computational grid {\citep{sig07}}.
The density structure of the disk is assumed to be of the form
\begin{equation} 
\label{rho_disk}
\begin{split}
\rho(R,z) & = \rho_o \left( \frac{R_{*}}{R} \right)^{n} e^{-(z/H)^2} \;,\\
            H(R)& = \sqrt{\frac{2 R^3 k T_{\rm{iso}}}{G M
\mu} } \, .
\end{split}
\end{equation}
Here $R_*$ is the stellar radius (see next paragraph). In the
second equation for the disk scale height, $H(R)$, $\mu$ is the
mean-molecular weight of the gas, and $T_{\rm iso}$ is an assumed
isothermal temperature used for the sole purpose of setting the density
scale height.  The models are constructed with the density parameters
set to $n=3$ and $\rho_{o}=5\,\times 10^{-11}$ g/cm$^3$. These density
parameters are kept constant.  The form of the disk density given by
Equation~\ref{rho_disk} follows from the assumption of a radial power-law
drop in the equatorial plane ($z=0$) coupled with the assumption that the
disk is in vertical hydrostatic equilibrium set by the $z$-component of
the star's gravitational acceleration. Note that Equation~\ref{rho_disk}
results in a flaring disk with $H\propto\,R^{3/2}$.

If the equatorial radius expands with rotation, this can
change the disk's density structure if the $R_*$ appearing in
Equation~\ref{rho_disk} is associated with the star's equatorial
radius. This is undesirable because we wish to investigate changes in
the disk temperatures resulting solely from the changing
photoionizing radiation field seen by the disk and not from an unintended
change to the underlying density structure of the disk. Unfortunately,
there is no ideal solution to this complication and several approaches
are outlined in Appendix~A.  In the notation of Appendix~A, all of the
calculations were computed with an unchanging grid in which the inner edge
of the disk is set to $3/2\,r_{\rm p}$. This approach preserves the disk density
structure and is best for temperature comparisons.  A physical grid, in
which the density structure of the non-rotating star is shifted outward
as $R_*$ increases, is more useful when calculating observables, and
this grid will be used in a subsequent paper on predicted H$\alpha$ line profiles
and infrared excesses.

\begin{table}[h]
\caption{Adopted stellar parameters \label{star_para}}
\begin{center}
\begin{tabular}{cccccc}
\hline
Spectral& Mass  & Polar Radius & Luminosity  & ${\omega}_{\rm{crit}}$ &
 $ v_{\rm{crit}}= \frac{3}{2}r_{\rm{p}}\,\omega_{\rm{crit}}$ \\
 Type & (\msun) &  (\rsun) & (\lsun) & (${\rm{s}^{-1}}$)  &  ($\rm km\,\rm s^{-1}$) \\ \hline\hline
  B0V & 17.5  & 7.40  & $3.98 \, {\rm x} \, 10^{4}$  & $ 7.10 \, {\rm x} \, 10^{-5}$ & $ 548$ \\
  B2V & 9.11  & 5.33  & $4.76 \, {\rm x} \, 10^{3}$  & $ 8.38 \, {\rm x} \, 10^{-5}$ & $ 466$ \\
  B3V & 7.60  & 4.80  & $2.58 \, {\rm x} \, 10^{3}$  & $ 8.95 \, {\rm x} \, 10^{-5}$ & $ 449$ \\
  B5V & 5.90  & 3.90  & $7.28 \, {\rm x} \, 10^{2}$  & $ 1.08 \, {\rm x} \, 10^{-4}$ & $ 439$ \\ \hline
\end{tabular}\\
\end{center}
\vspace{0.1in}
{\it Notes:} Stellar parameters adopted from \citet{cox2000}.
\end{table}

\begin{table}[htp]
\caption[]{Adopted rotation rates and example equatorial speeds. \label{v_to_w}}
\begin{center}
\begin{tabular}{cccccc }
\hline
 $v_{\rm{frac}}$  & ${\omega}_{\rm{frac}}$ & ${r_{{\rm{eq}}}}/{r_{\rm{p}}}$ & ${v_{\rm{eq}}}$ B0 
& ${v_{\rm{eq}}}$ B3 & ${v_{\rm{eq}}}$ B5 \\ 
            &            &       & ($\rm km\,\rm s^{-1}$) & ($\rm km\,\rm s^{-1}$) & ($\rm km\,\rm s^{-1}$) \\ \hline\hline
   0.000    &    0.000  &   1   &   0   &   0   &  0  \\ 
   0.001    &    0.002  &  1.00 & 0.55 & 0.45 & 0.40\\ 
   0.200    &    0.296  &  1.01 &  110  &  90 & 80 \\ 
   0.400    &    0.568  &  1.06 &  219  &  180  & 160 \\ 
   0.600    &    0.792  &  1.14 &  329  &  269  & 241 \\ 
   0.800    &    0.944  &  1.27 &  438  &  359  & 321 \\ 
   0.950    &    0.996  &  1.43 &  521  &  427  & 381 \\ 
   0.990    &    0.999  &  1.49 &  544  &  445  & 397 \\ \hline 
\end{tabular}
\end{center}
\end{table}

The photoionizing radiation field incident on the disk was
calculated as follows: first, von Zeipel's constant is found by using
Equation~\ref{C_scale} to rescale an interpolated value of $C_\omega$
from \cite{col66a} to match the stellar parameters.  Next, for each point
in the disk, the visible region of the stellar surface is determined,
and a grid is constructed across this surface.  At each point in this
surface grid, the local temperature, local gravity, and the viewing
angle are found.  Finally, the specific intensity, $I_{\nu}\left(
T_{\rm{eff}}, g\right)$, for each point is scaled from the \citet{kur93}
model atmosphere closest to the effective temperature and gravity at
each location, $I^{K}_{\nu}\left( T_{\rm{best}}, g_{\rm{best}}\right)$,
to the required temperature using a blackbody function,
\begin{eqnarray}
\label{I_rescale}
I_{\nu}\left(T_{\rm{eff}},g \right)=I^{\rm{K}}_{\nu}\left( T_{\rm{best}}, g_{\rm{best}}\right) \frac{B_{\nu} \left( T_{\rm{eff}}\right) }{B_{\nu} \left( T_{\rm{best}} \right)}\,.
\end{eqnarray}
\noindent
Here $T_{\rm{eff}}$ is the local effective temperature of the stellar
surface,
and $T_{\rm best}$ and $g_{\rm best}$ are the parameters of the closest matching
Kurucz model.
The available Kurucz grid of model atmosphere intensities was
interpolated to a finer grid of spacing of $\Delta T_{\rm{eff}} = 200
{\rm K} $ and $\Delta \log (g) = 0.10$. Equation~\ref{I_rescale} was used
to correct the intensity only in these very narrow temperature bins.

Most of the calculations performed in this work implement the
above description of gravitational darkening and are referred to as
full gravitational darkening (FGD) calculations. All of the models
presented in Section~\ref{results} are FGD calculations. However in
Section~\ref{SGD/PSD}, two simplified approaches are described which
isolate one of the two main effects of gravitational darkening: spherical
gravitational darkening (SGD), accounting only for the temperature
change across the stellar surface, and pure shape distortion (PSD),
accounting only for the changing the shape of the star.

\section{Results for Gravitationally-Darkened Be Star Disks}
\label{results}

\subsection{Energy reaching the Circumstellar Disk}
\label{results1}

The stellar radiation field is modified by gravitational darkening, and
this directly affects the disk temperatures through the photoionization
rates.  Figure~{\ref{Energy_input} shows the total photoionizing mean
intensity at each location in the computation grid for a B2 star both
with and without a disk, and with and without rotation.  The mean
intensity ratios between the various cases are also shown. Consider
the left-hand panels A and B in which there is no disk material.  In the
rotating case ($v_{\rm frac}=0.99$), less radiation from the poles
reaches the equatorial regions of the disk, and this results in the hour
glass shape in the mean intensity of panel~B.  In the bottom-left panel
which gives the ratio of the rotating to non-rotating case, the stellar
radiation field reaching the disk has been reduced by approximately
50\% compared to the non-rotating star. However, for regions close
to the star and near the equatorial plane, there can be reductions as
great as 90\%.  Now consider the right-hand panels C and D in which the
opacity and emissivity of the disk gas have been included. Again the
cases of rotation at $v_{\rm frac}=0.99$ and no rotation are compared.
Most of the radiation reaching various positions within the disk is
re-processed by the circumstellar gas between a particular location and
the star. In general, the total irradiance reaching the shielded parts of
the disk is reduced down with gravitational darkening by about 50\%.  However,
there are locations within the mid-regions of the inner disk where the
energy actually increases with rotation (see Figure~{\ref{Energy_input})
due to changes in the disk's opacity.

\begin{figure}
\epsscale{.75}
\plotone{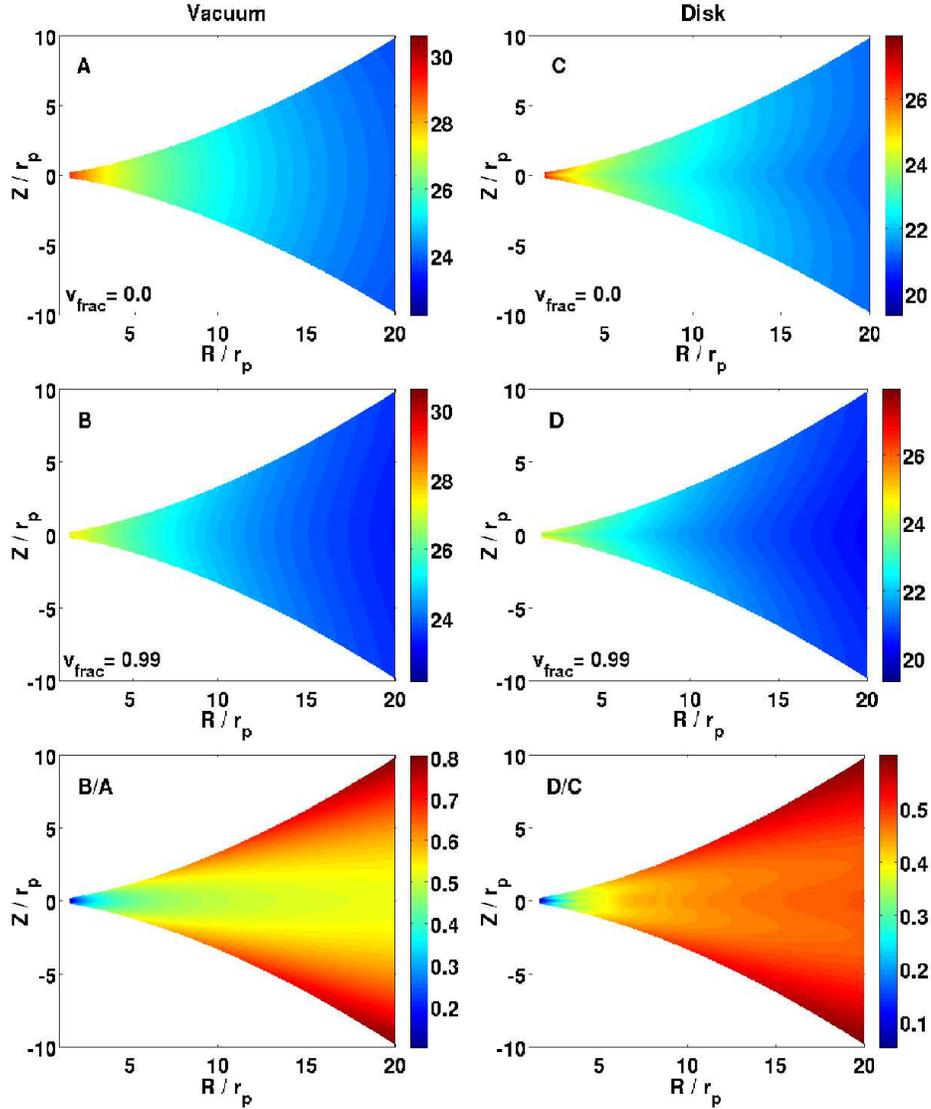}
\caption{Left panels A and B show the log of the mean intensity at each
location in the computational grid without any disk material present
for a star with $v_{\rm{frac}}=0$ (panel A) and $v_{\rm{frac}}=0.99$
(panel B). The ratio of the mean intensity for the rotating case with the
non-rotating case is given in the bottom left panel (labelled B/A).  
The right panels C and D show the mean
intensity at each location in the disk including the opacity effects
of the disk material for a star with $v_{\rm{frac}}=0$ (panel D) and
$v_{\rm{frac}}=0.99$ (panel E). The ratio of the shielded
rotating case with the non-rotating case is shown
in bottom right panel (labelled D/C).
\label{Energy_input}}
\end{figure}

\subsection{Temperatures in the Circumstellar Disk}
\label{results2}

Any change in the energy received from the star produces a
change in the radiative equilibrium temperature. The
changes produced by increasing rotation on the general temperature
structure of the disk are illustrated in Figures~\ref{global_temp_short}
through \ref{global_temp_short_B5} for the spectral types given in
Table~\ref{star_para}.  Shown in all figures are four disk temperature
diagnostics: the maximum temperature, the minimum temperature, the
density-weighted temperature, defined as
\begin{equation}
\label{T_mass_ave}
\overline{T}_{\rho} = \frac{1}{M_{\rm{disk}}}\,\int T(R,z)\,\rho(R,z)\,dV \;,
\end{equation}
and the volume-averaged temperature, defined as
\begin{equation}
\label{T_vol_ave}
\overline{T}_{\rm{V}} = \frac{1}{V_{\rm{disk}}}\,\int T(R,z)\,dV \;.
\end{equation}
In order to avoid numerical effects, the maximum and minimum
temperatures are averages over the twenty hottest and twenty
coolest disk locations, respectively.  In Figures~\ref{global_temp_short} through
\ref{global_temp_short_B5}, all of these temperature diagnostics
are plotted as a function of $v_{\rm{frac}}$ and $\omega_{\rm{frac}}$.
Because their behaviour depends somewhat on spectral type,
the results for each model are discussed separately.

\begin{figure}
\epsscale{.60}
\plotone{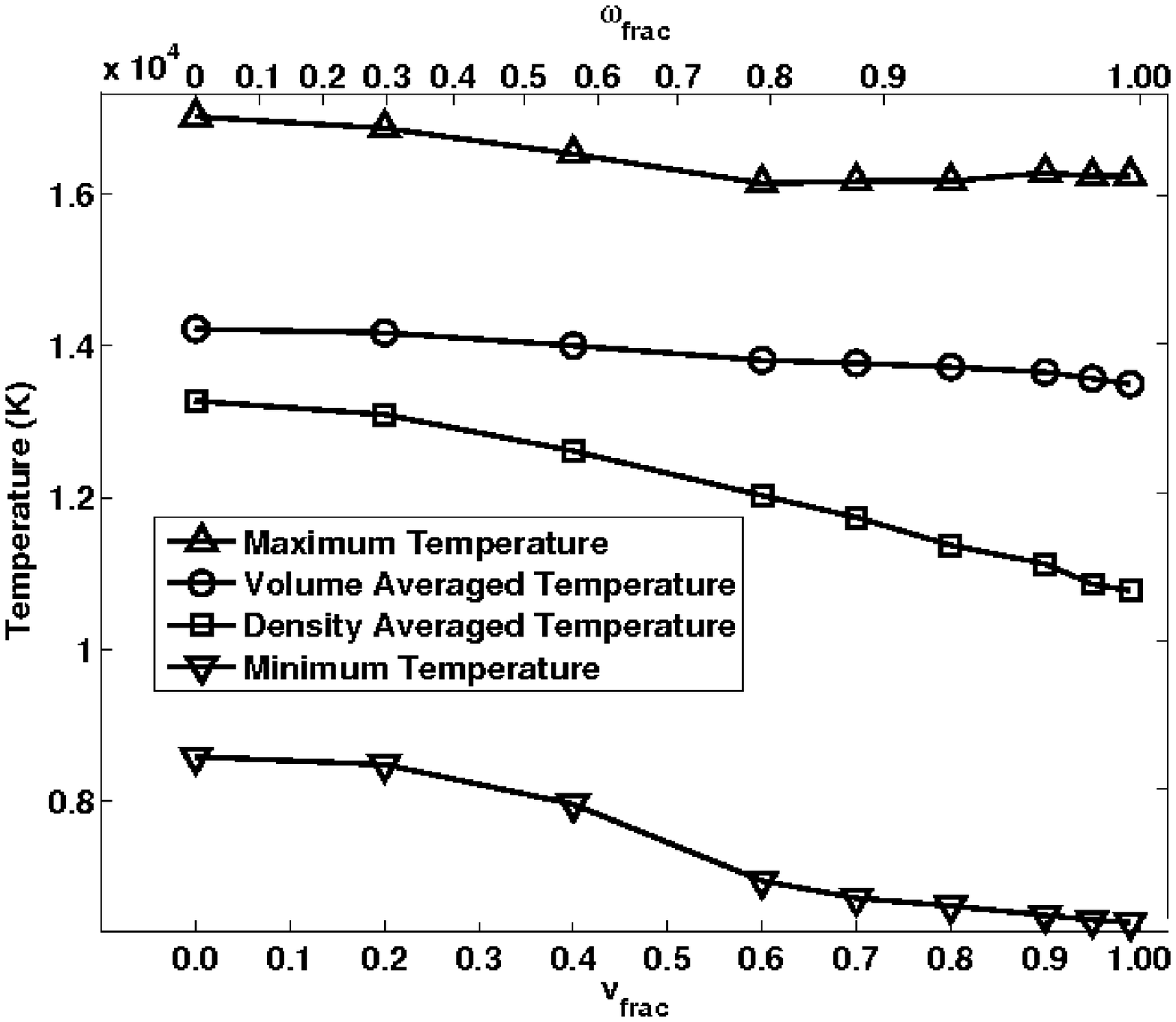}
\caption{Change in disk temperature diagnostics with increasing rotation 
for the B0 model. \label{global_temp_short}}
\end{figure}

\begin{figure}
\epsscale{.60}
\plotone{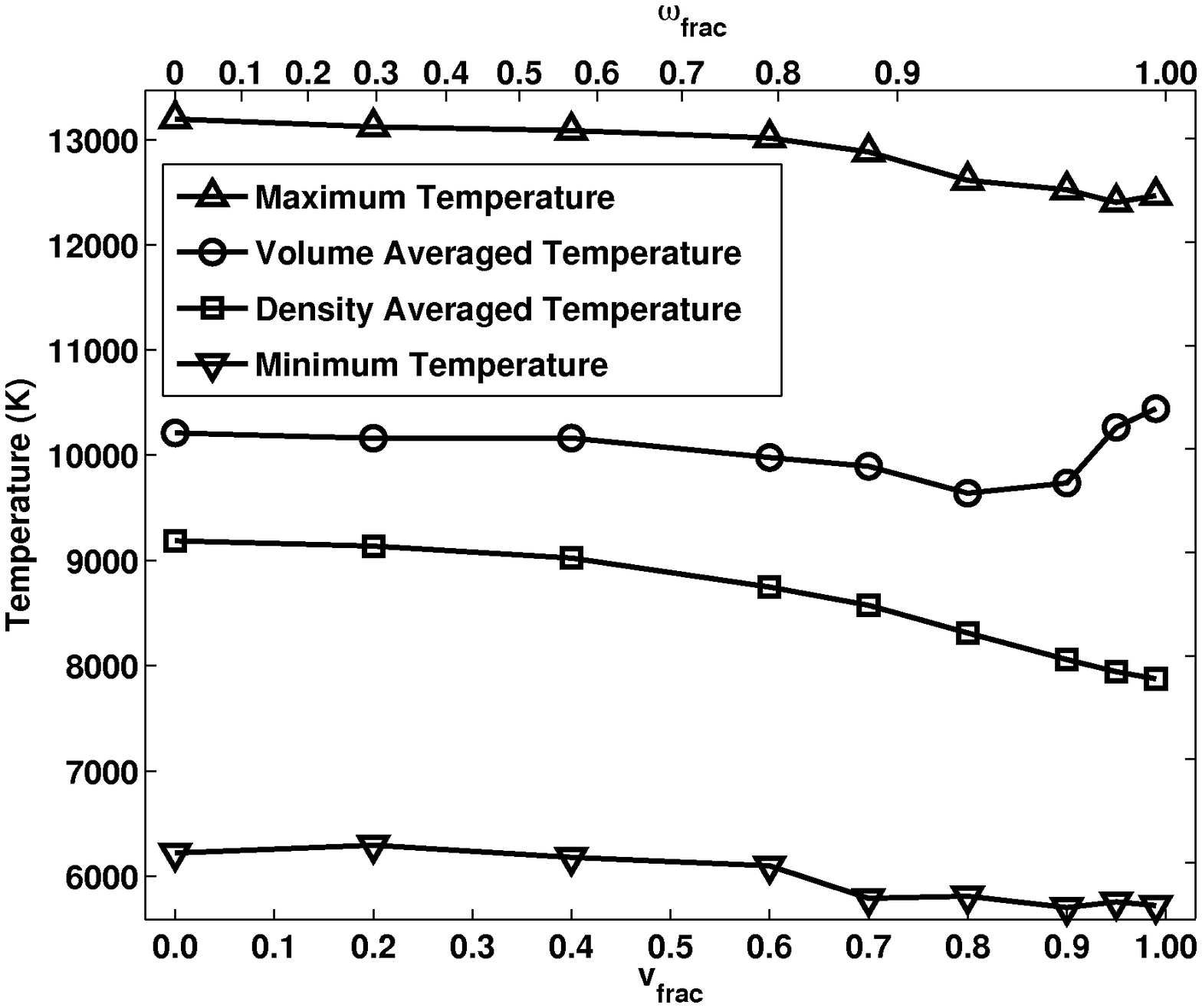}
\caption{Same as Figure~\ref{global_temp_short} for a B2 star. 
\label{global_temp_short_B2}}
\end{figure}

\begin{figure}
\epsscale{.60}
\plotone{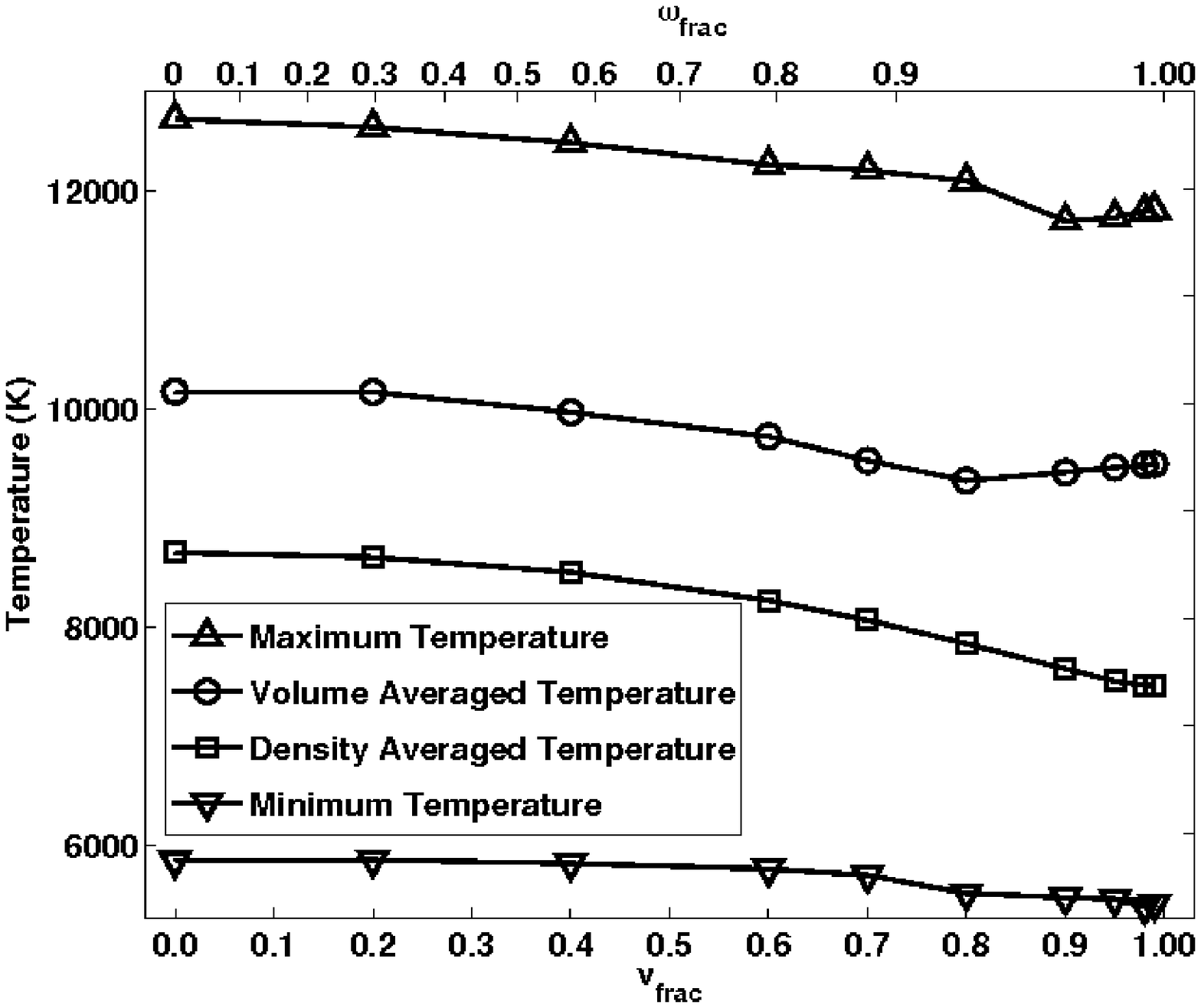}
\caption{Same as Figure~\ref{global_temp_short} for a B3 star. 
\label{global_temp_short_B3}}
\end{figure}

\begin{figure}
\epsscale{.60}
\plotone{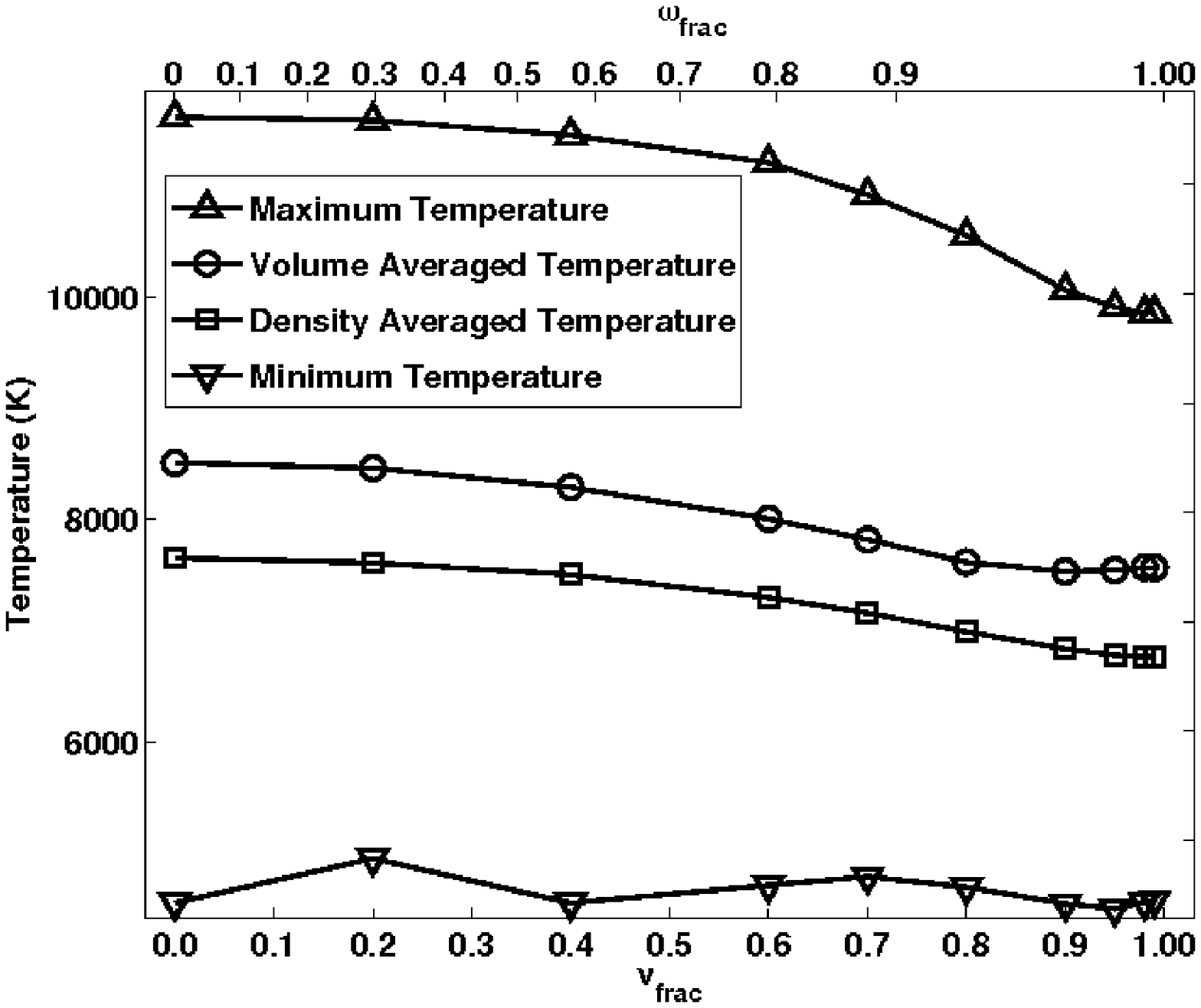}
\caption{Same as Figure~\ref{global_temp_short} for a B5 star. 
\label{global_temp_short_B5}}
\end{figure}

%

For the B0 model, seen in Figure \ref{global_temp_short}, both the
density and volumed-averaged disk temperatures decrease steadily
with increasing rotation. Small declines in the density-weighted,
maximum, and minimum disk temperatures are apparent even at
$v_{\rm{frac}}\approx\,0.4$.  By critical rotation, the density-weighted
temperature has fallen to just under $11,000\,$K, a decline of $2500\,$K
compared to the non-rotating case. The decline is not as large in the
volumed-average temperature, and this is likely due to the influence
of the hotter stellar pole that develops for increased rotation (see
Figure~\ref{Temperature_profile}); this is also reflected in the maximum
and minimum disk temperatures: while the minimum temperature has a steep
decline, the maximum temperature is much flatter and actually reaches
a shallow minimum at an intermediate rotation rate.

For the B2 model, shown in Figure~\ref{global_temp_short_B2}, all
temperatures initially decline slowly. The density-weighted disk
temperature falls from over $9000\,$K in the non-rotating case to under
$8000\,$K by critical rotation. However, unlike the previous case, the
volume-averaged temperature reaches a minimum near $v_{\rm{frac}}\approx
0.8$ and then begins to increase for extremely rapid rotation.  The lack
of an increase in the density-weighted temperature indicates that the
additional heating occurs mainly in the upper disk, where there is less
gas, and the heating is likely due to the effect of the hot stellar pole.
We also note that minimum temperature in the disk, $\approx\,6000\,$K,
is approximately constant with rotation, with only a small decline for
$v_{\rm{frac}}\ge\,0.7$.

Spectral type B3 is shown in Figure~\ref{global_temp_short_B3}. Its
behaviour is very similar to the B2 case above, although the
temperature rise in the volume-weighted average temperature
above $v_{\rm{frac}}\ge\,0.8$ is not as large as in the B2
case. Again the minimum disk temperature is around $6000\,$K and
is not strongly affected by rotation.  For B5, shown in
Figure~\ref{global_temp_short_B5}, both the density and volume-averaged
temperatures show a similar decline with rotation and there is no rise
in the volume-averaged temperature above $v_{\rm{frac}}\!\ge\!0.9$.
By this spectral type, the minimum disk temperature has fallen to
$\approx\,5000\,$K and shows no clear trend with rotation. The low
temperatures reached in this model have implications for later spectral
types that we shall now discuss.

As noted in the introduction to Section~\ref{cals}, we have not considered
a model of spectral type B8 from the VL (very late) spectral bin of
\citet{cra05}.  Our test calculations have indicated that very cool disk
temperatures are reached in this model, particularly for near critical
rotation, and these may fall outside of the domain of applicability of
the current version of {\sc bedisk}. In particularly while the global
density-weighted disk temperature still exceeds $6000\,$K for a B8 model,
there is a significant volume of the disk which falls well below $5000\,$K by
$v_{\rm{frac}}=0.99$. While our calculations include abundant metals
with low ionization potentials (such as Mg, Ca, and Fe) that provide
sources of free electrons at low temperatures, molecule formation is
not included. In addition for such cool disks, the treatment of the
diffuse radiation field generated within the disk may require a more
careful treatment. For these reasons, we have chosen not to include B8
models in this work. We note that in the analysis of \citet{cra05}, the
four bins we do account for include the full range of inferred threshold
rotation rates, from $v_{\rm frac}\!\approx\!0.6$ for early spectral types
B0--B2 to $v_{\rm frac}\!\approx\!1.0$ for the late spectral type B5.

The effect that near critical rotation ($v_{\rm{frac}}=0.99$) has
on the global, density-weighted disk temperature is summarized in
Table~\ref{tab:trho}. Compared to non-rotating models, near critical
rotating models are between 15 and 20\% cooler with the largest difference
occurring for the earliest spectral type considered (B0).

\begin{figure}
\epsscale{.80}
\plotone{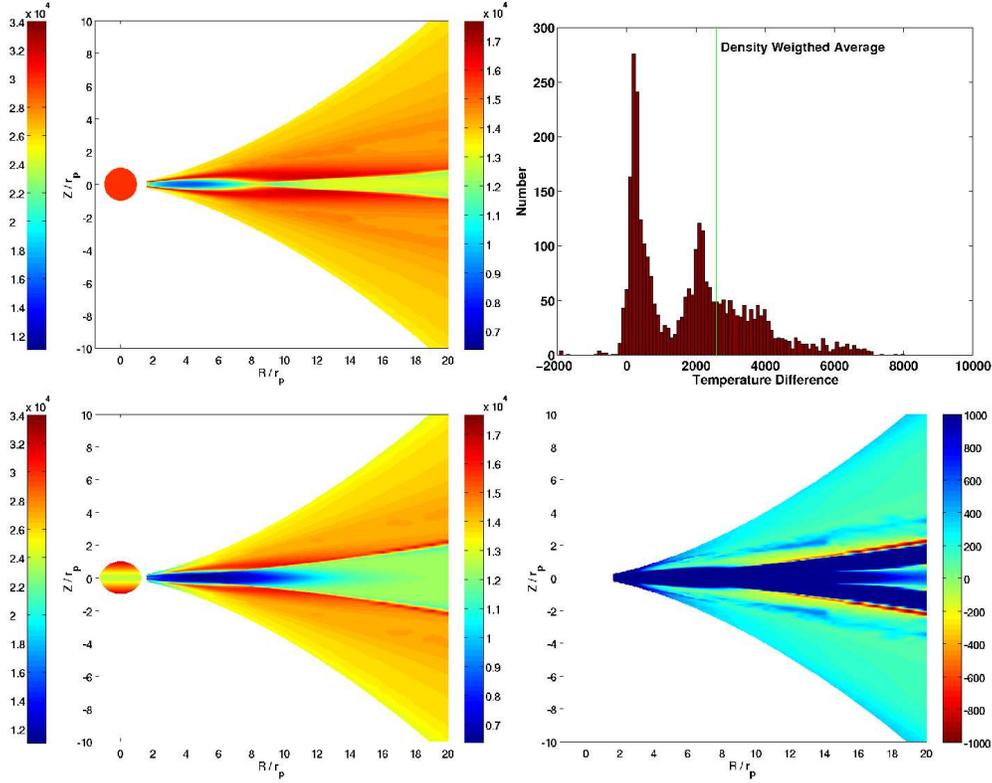}
\caption{Temperatures for both a non-rotating B0
star and its circumstellar disk (upper left) and a B0 star
rotating at $v_{\rm{frac}}=0.95$ with an identical disk (lower
left). In each of these panels, the colour bar on the left is for the
star and the colour bar on the right, for the disk.
The lower right panel shows the temperature differences in the
disk. Positive differences mean that the non-rotating star is hotter. The
upper-right panel shows a histogram of the temperature differences.
\label{Temperature_panels_b0}} 
\end{figure}

\begin{figure}
\epsscale{.80}
\plotone{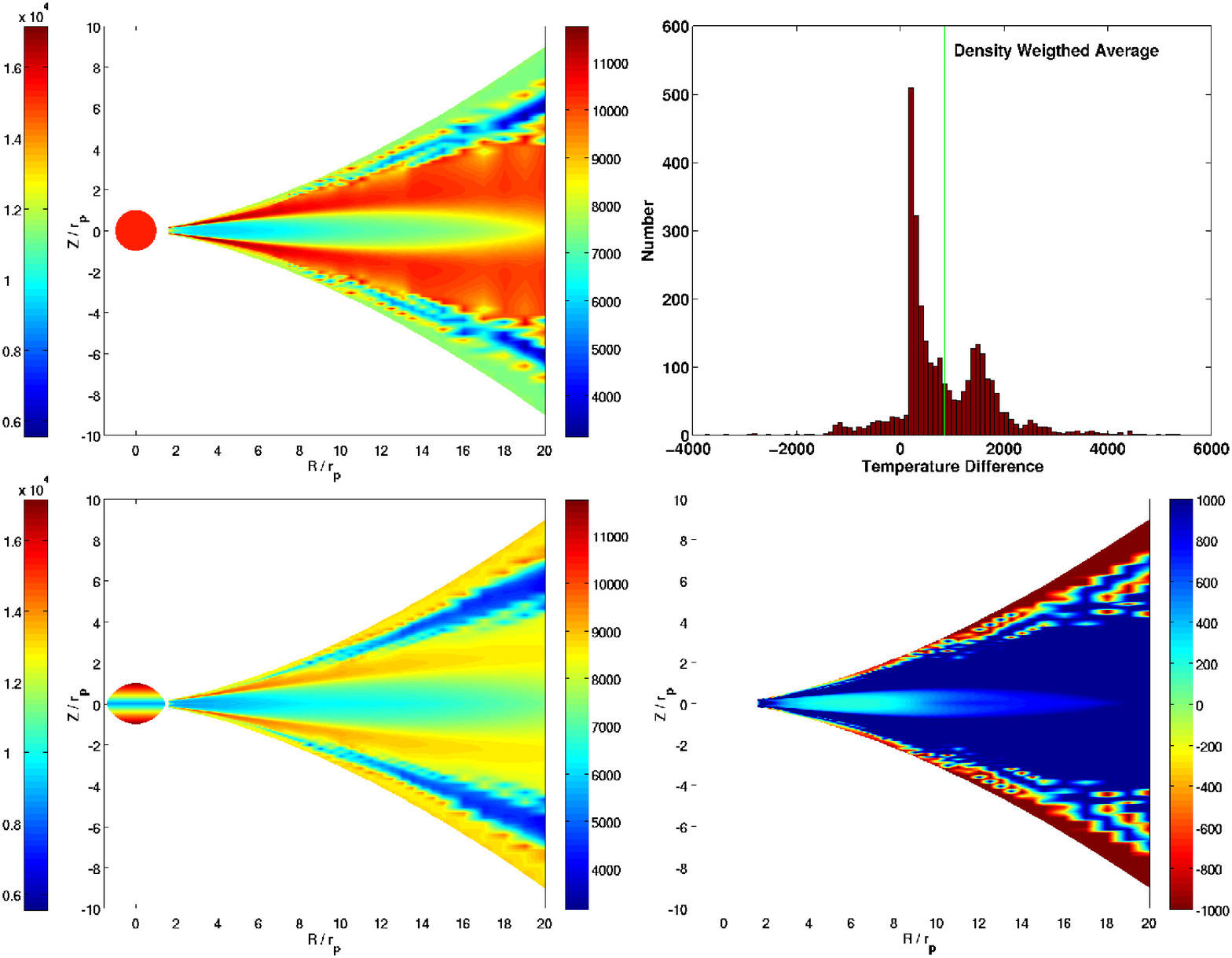}
\caption{Same as Figure~\ref{Temperature_panels_b0} for a B5 star.
\label{Temperature_panels_b5}}
\end{figure}

\begin{deluxetable}{crrc}
\tablewidth{0pt}
\tablecaption{Change in density-weighted average disk temperature
for near-critical rotation.\label{tab:trho}}
\tablehead{
\colhead{Spectral Type} & \multicolumn{2}{c}{\hrulefill\,$\overline{T}_{\rho}(K)$\,\hrulefill} 
& \colhead{\% Change} \\
  ~                     & \colhead{$v_{\rm frac}=0.00$} & \colhead{$v_{\rm frac}=0.99$} & ~
}
\startdata
B0      & 13280 & 10780 & -19\\
B2      &  9190 & 7870  & -14\\
B3      &  8690 & 7460  & -14\\
B5      &  7660 & 6760  & -12\\
\enddata
\end{deluxetable}

Next we discuss how the detailed temperature structure of a Be star disk,
$T(R,z)$, changes with rotation.  In general, Be star circumstellar
disks are highly non-uniform in temperature as disks of sufficient
density develop a cool, inner zone surrounded by a larger warm region
\citep{mil98,sig09b}.  Between these two regions is the hottest part of
the disk, forming a narrow sheath.  Figures~\ref{Temperature_panels_b0}
and \ref{Temperature_panels_b5} show the effect of rotation on
the temperature of disks surrounding stars of spectral types B0
and B5 respectively. In each plot, temperature distributions for
two models are shown, one with $v_{\rm{frac}}=0.0$ and the other
with $v_{\rm{frac}}=0.95$.  Also shown is the temperature difference
between the rotating and non-rotating models and a histogram of these
differences. The value of $v_{\rm{frac}}=0.95$ was chosen for the
rotating model because this is approximately the rotation rate required
for a velocity perturbation on the order of the local sound speed in the
star's photosphere to be sufficient to feed material into a rotationally
supported disk.

For the B0 model, shown in Figure~\ref{Temperature_panels_b0}, the
most dramatic changes in temperature occur because of the expansion
of the cool, inner zone to larger radii with increased rotation.
In the non-rotating case, this cool region does not extend past
$\approx7\,R_*$ while in the $v_{\rm{frac}}=0.95$ model, it extends
well past $10\,R_*$.  This results in some computational grid
points that are more than $4,000\,$K cooler in the rotating
model (compared to the $2500\,$K difference in the density-weighted
average) as illustrated by the histogram of temperature differences in
Figure~\ref{Temperature_panels_b0}.  However, there are also temperature
changes in the hot sheaths above and below the inner cool zone close to
the star. For the $v_{\rm{frac}}=0.95$ model, these hot sheaths occur
only beyond $\approx 3\,R_*$ whereas in the non-rotating case, these
sheaths extend to the inner edge of the disk.

Despite these changes, the temperature in the equatorial plane of
the cool, inner zone does not drop dramatically with rotation, likely
due to the high optical depths along the rays back to the star. This
explains why the density-weighted average temperature of the disk
shows only a modest decrease compared to the non-rotating case (see
Figure~\ref{global_temp_short}). Interestingly, the rotating model
is actually hotter than the non-rotating model in a narrow region a
few scale heights above the equatorial plane for $R>10 R_*$, although
the number of computational grid points involved is quite small. The
optically thin gas above and below the plane of the disk shows only a very
small decrease in temperature and thus the volume-averaged temperature
of Figure~\ref{global_temp_short} is essentially unaffected by rapid
rotation.

Figure~\ref{Temperature_panels_b0} clearly illustrates that the
temperature distribution surrounding a rotating star may differ
dramatically from the non-rotating case, even when global measures of
disk temperature such as $\overline{T}_{\rho}$ and $\overline{T}_{V}$
are similar. From the point of view of computing the strength of emission
lines or the infrared excess, the increased extent of the inner
cool zone and the changes to the temperature of the hot sheaths above
and below this zone have the potential to produce large changes in
these observational diagnostics.  Such changes will be discussed in a
subsequent paper.

Figure~\ref{Temperature_panels_b5} shows a similar comparison between
a non-rotating model and one rotating at $v_{\rm{frac}}=0.95$
for spectral type B5. As in the B0 case, the main effect is the
extension of the inner, cool zone to larger radii. However the B5
model also has a significant extension of this cool zone to larger
distances above (and below) the plane of the disk. As these regions
occupy a significant volume, the volume-weighted disk temperature is
significantly decreased (unlike the previous case of a B0 star). As
shown in Figure~\ref{global_temp_short_B5}, both the density and
volumed-weighted disk temperatures show a similar drop with rotation.

While most disk locations are, as before, cooler, there is a much larger
fraction of disk locations that are hotter in the B5 rotating model than
in the previous B0 case. Most of these hotter regions are associated
with a cool zone at large radii, $R>10\,R_*$, far above (and below) the
equatorial plane of the disk. In these regions, the lines of sight back
to the central star have significant optical depths whereas the
optical depths (in the $z$-direction) to the nearest disk edge
have become less than unity. Hence heating via photoionization is
suppressed whereas cooling via escaping line radiation becomes effective
and there is a reduction in temperature due to this cooling. This low
temperature zone disappears for higher $z$ because the optical depths
back to the central star eventually drop and the heating rate increases.
The exact location and detailed form of the zones is influenced by
rotation and some locations with significantly hotter temperatures
are predicted.  It should be kept in mind that at large radii and far
above or below the plane of the disk, the gas densities are very low,
and the detailed structure of these cool zones will have little effect
on observable diagnostics such as emission line strengths, polarization,
or infrared excess.

\begin{figure}
\epsscale{.60}
\plotone{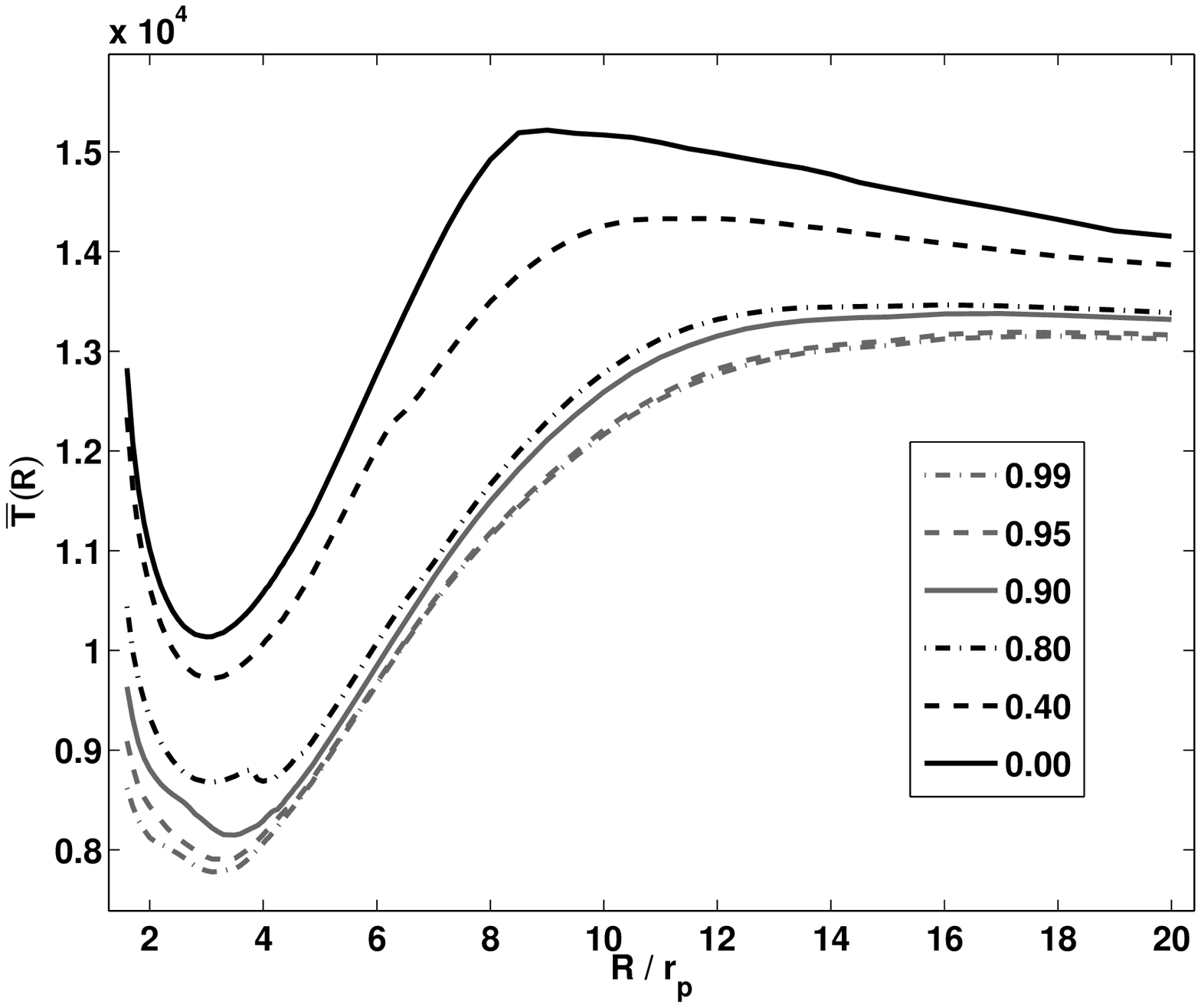}
\caption{Variation of the vertically-averaged, density-weighted temperature with 
disk radius for various stellar rotation rates for the B0 model. \label{temp_mid}}
\end{figure}

\begin{figure}
\epsscale{.60}
\plotone{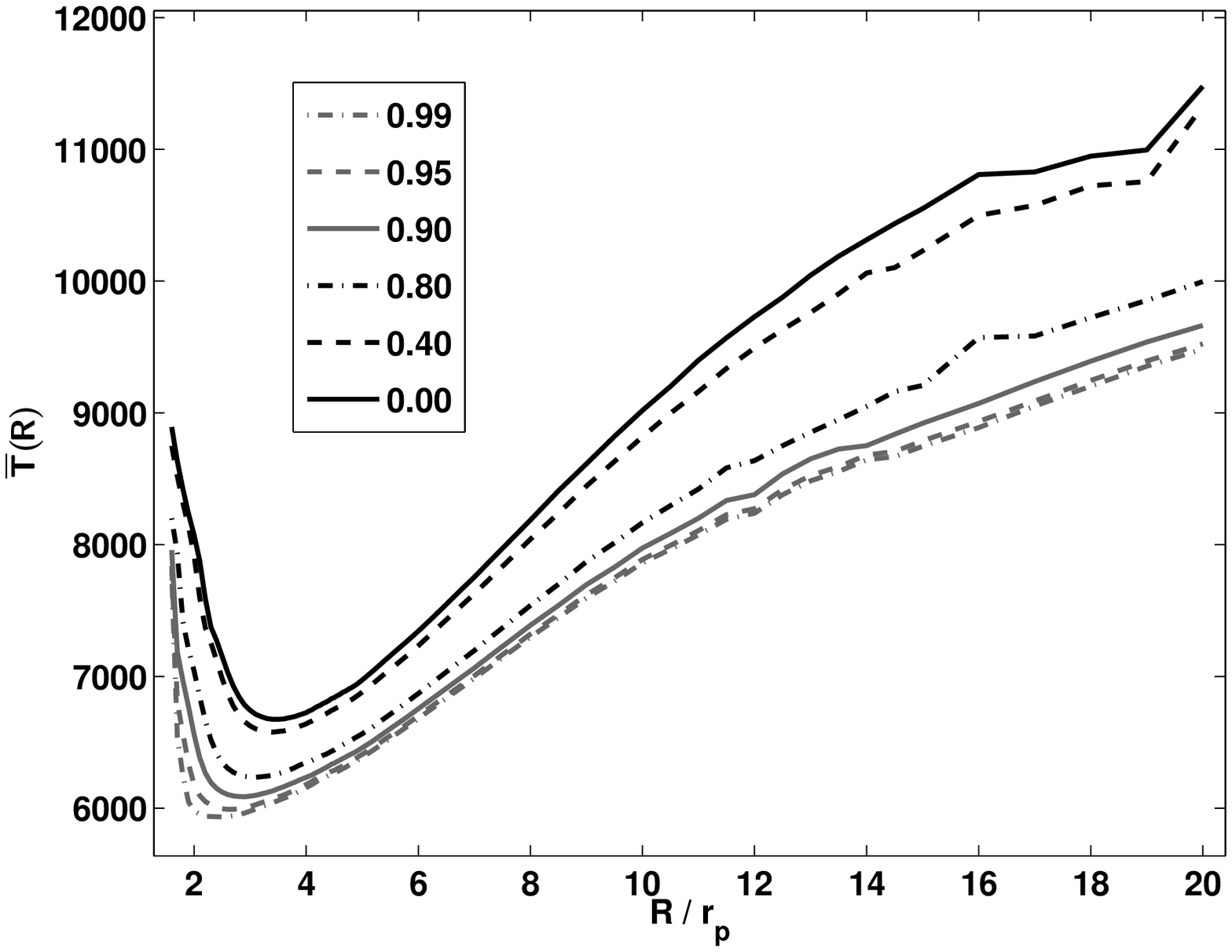}
\caption{Same as Figure~\ref{temp_mid} for the B3 model. \label{temp_mid3}}
\end{figure}

\begin{figure}
\epsscale{.60}
\plotone{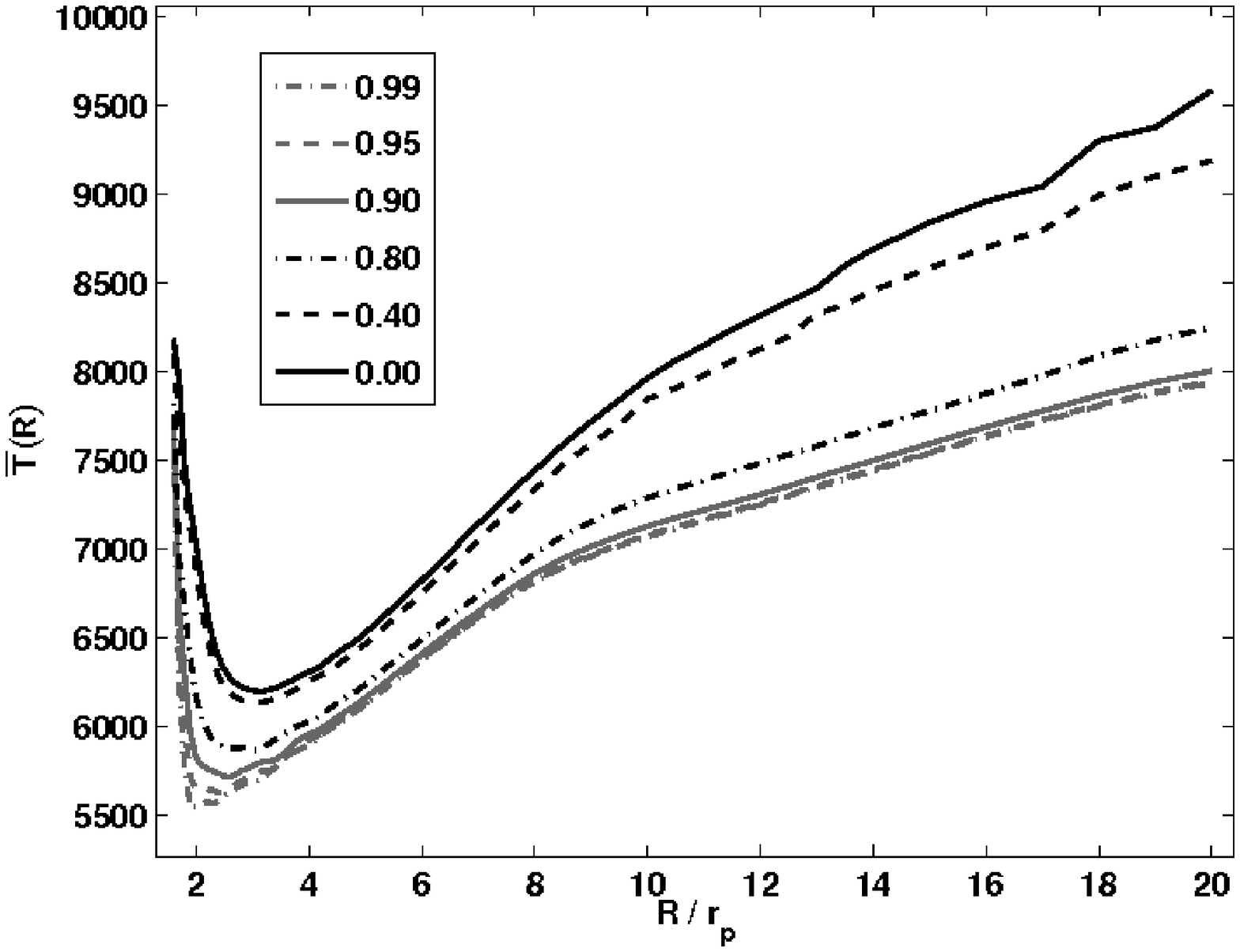}
\caption{Same as Figure~\ref{temp_mid} for the B5 model. \label{temp_mid5}}
\end{figure}

While the detailed $T(R,z)$ comparisons of the previous
paragraphs are instructive, it is difficult to see general trends
over a wide range of rotation rotates. For this reason, it is useful
to have temperature diagnostics that are intermediate between the
disk-averaged, global measures of Figures~\ref{global_temp_short} to
\ref{global_temp_short_B5} and the detailed temperature distributions
of Figures~\ref{Temperature_panels_b0} and \ref{Temperature_panels_b5}.
To this end, we have performed both vertical $(z)$ and radial $(R)$
temperature averages over the disks of three stellar models (B0, B3,
and B5) and considered how these averages are affected by a wide range
of rotation rates.

The vertical temperature averages (yielding an average disk temperature
as a function of distance from the star) were computed via
\begin{equation} 
\label{R_average_definition}
\overline{T}(R) = \frac{\int^{z_{\rm{max}}}_{0} T(R,z)\rho(R,z)\,dz}{\int^{z_{\rm{max}}}_{0} \rho(R,z)\,dz}\;,
\end{equation} 
and the change in vertically-averaged, density-weighted temperature
as a function of radius is shown in Figures~{\ref{temp_mid}} through
{\ref{temp_mid5}} for three stellar models, B0, B3, and B5. Five rotation
rates, $v_{\rm frac}=0.00$, $0.40$, $0.80$, $0.90$, $0.95$ and $0.99$, were
considered.  The behaviour of the B2 model of Table~\ref{star_para}, not
shown for brevity, is similar to the B3 case.  In all cases considered,
the $v_{\rm frac}=0.40$ models show vertically averaged temperatures that
are very close to the non-rotating case while significant differences
develop by $v_{\rm frac}=0.80$.

These vertical averages show that the region next to the central star
is hottest, followed by a rapid decrease towards a minimum temperature
near $\approx4\;R_*$ (for B0) to $\approx6\;R_*$ (for B5). Further out,
the temperature increases again towards the outer region of the disk. In
the B0 model, the temperature then plateaus, whereas in the B3 and B5
models the temperature is still increasing by $R\approx\,20\,R_*$. From
these figures, it can be seen that increasing rotation tends to: (1)
decrease the average temperature at all radii, (2) move the location
of the temperature minimum in the cool zone inward (although not in
the B0 model), and (3) broaden the extent of the cool, inner zone.
Rotation also tends to flatten the temperature increase occurring in
the outer regions of the disk.


\begin{figure}
\epsscale{.60}
\plotone{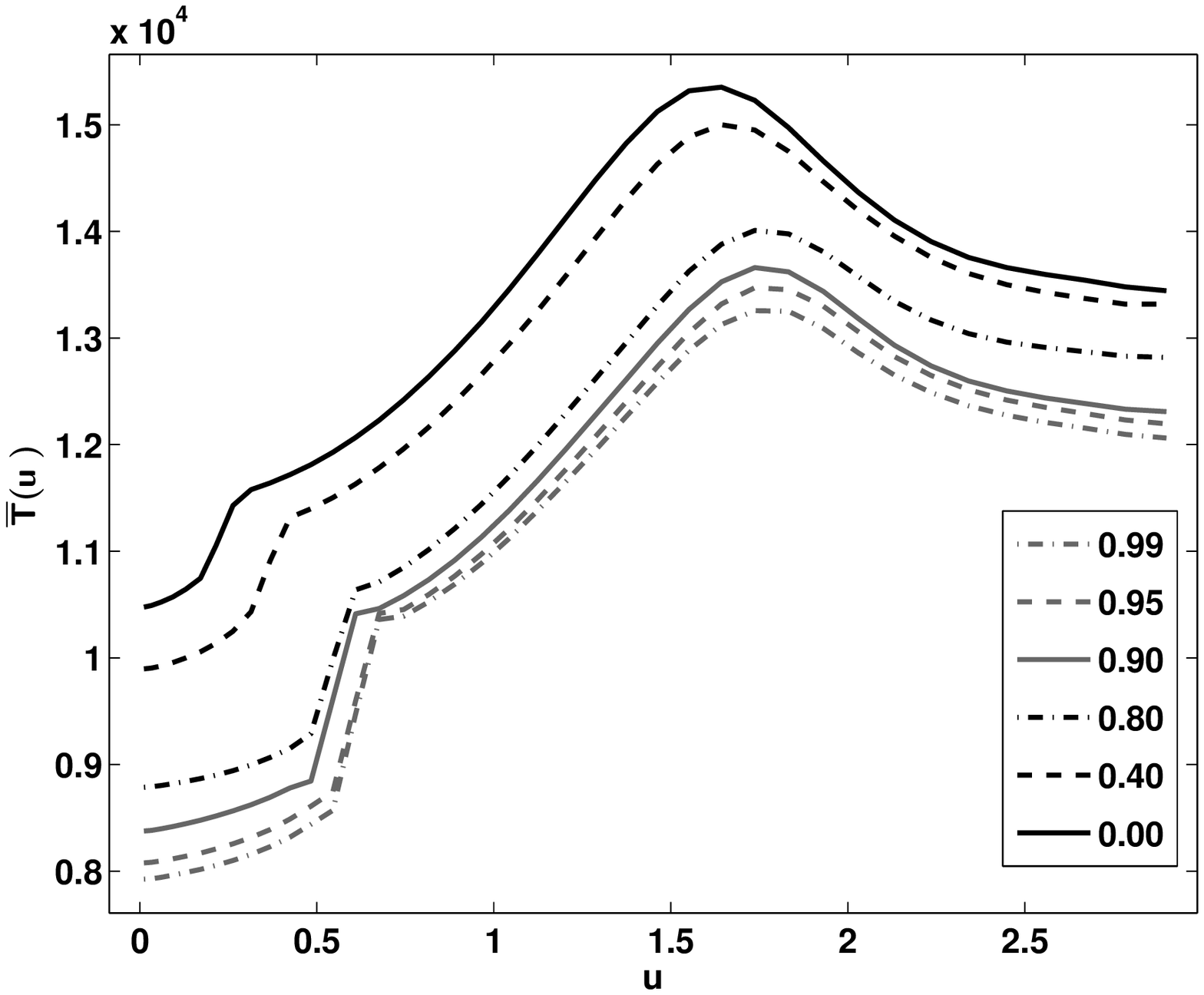}
\caption{Variation of the radially averaged temperature profile in the
vertical direction for the B0 model.  Here the gas scale height is the
variable $u$ defined by $u\equiv z/H(R)$. \label{temp_z}} 
\end{figure}

\begin{figure}
\epsscale{.60}
\plotone{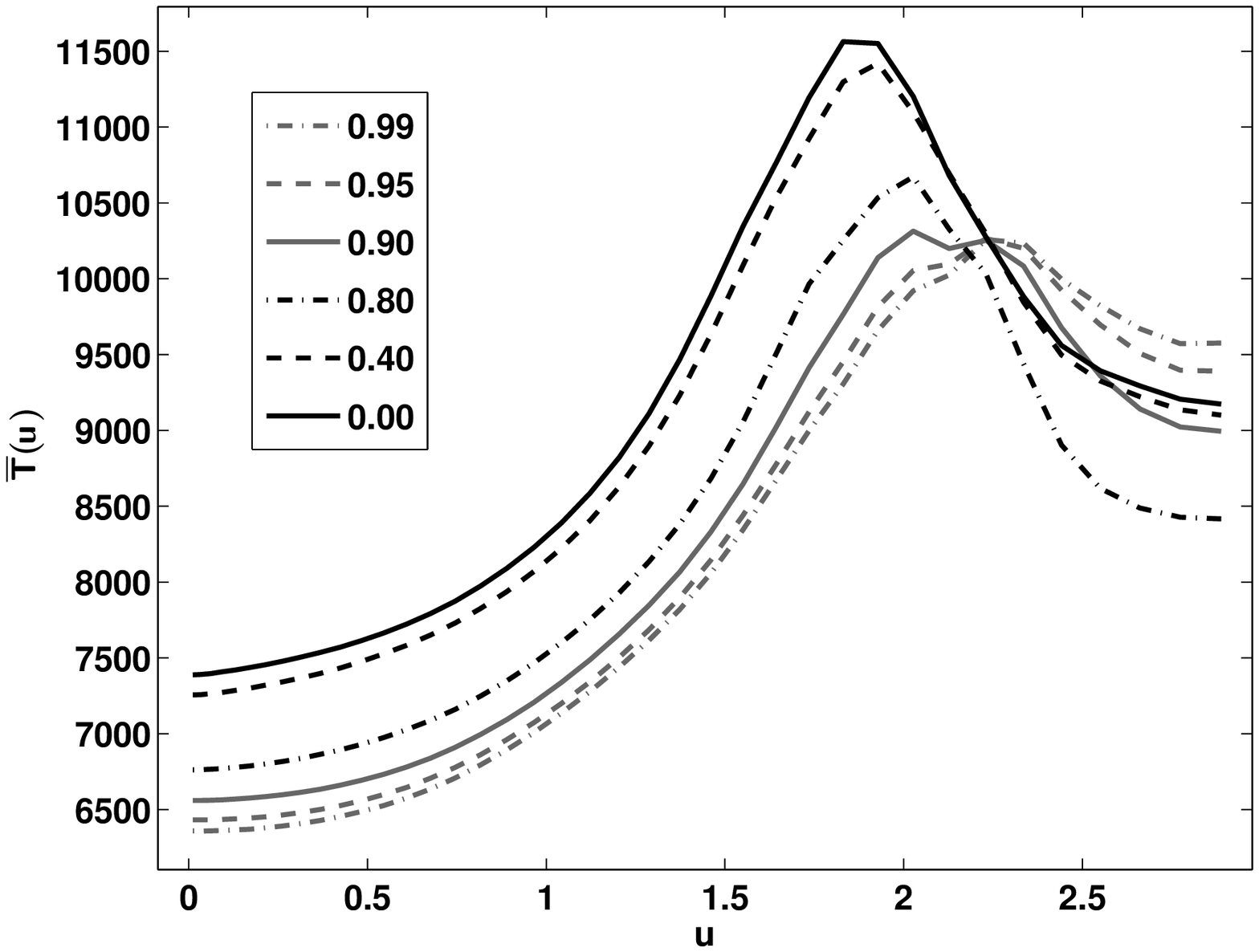}
\caption{Same as Figure~\ref{temp_z} for the B3 model. \label{temp_z3}}
\end{figure}

\begin{figure}
\epsscale{.60}
\plotone{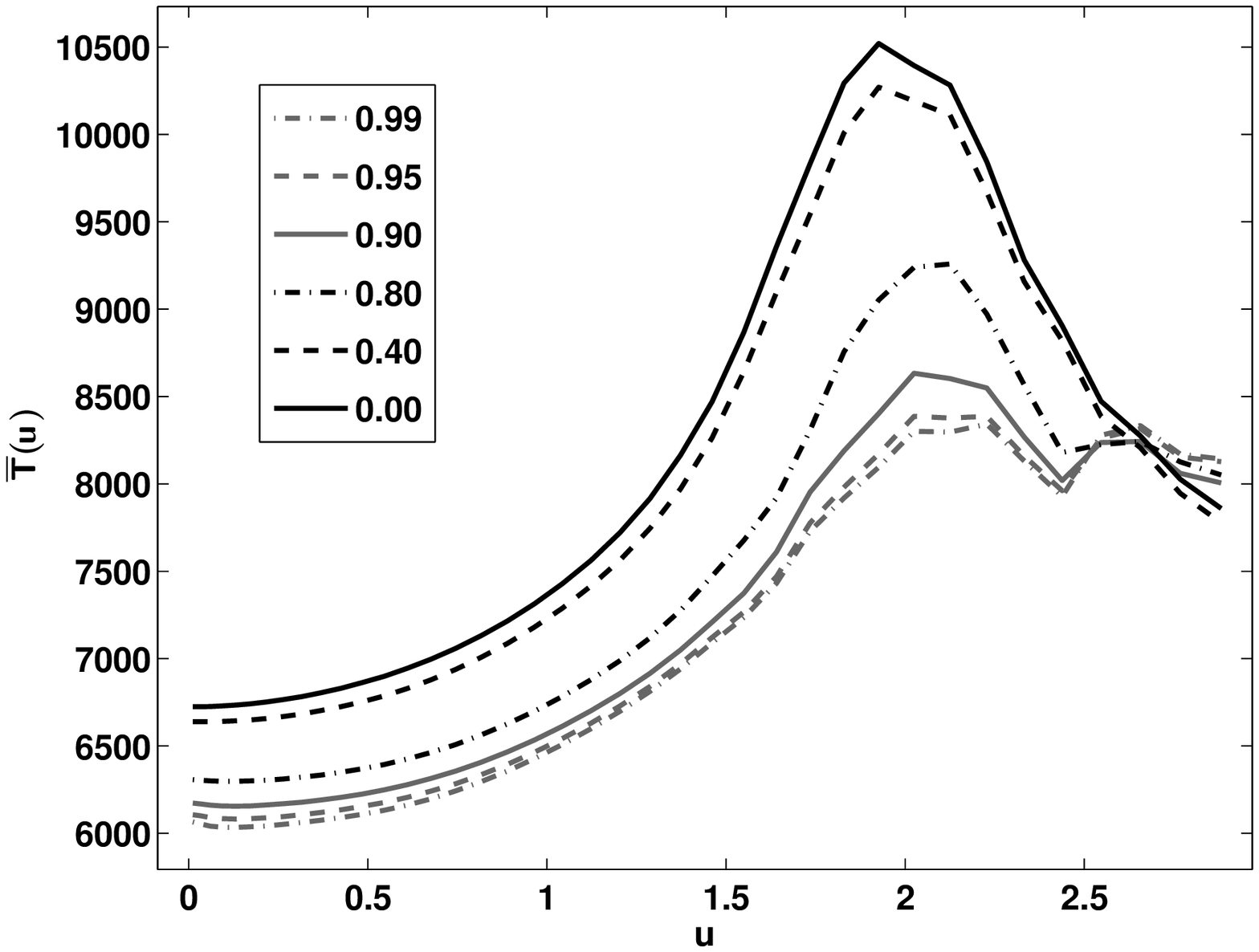}
\caption{Same as Figure~\ref{temp_z} for the B5 model. \label{temp_z5}}
\end{figure}

Turning to the  vertical $(z)$ structure of the disks, we define a
density-weighted, radially-averaged temperature (yielding an average
temperature as a function of height above the equatorial plane) as
\begin{equation} 
\label{z_average_definition}
\overline{T}(u) = \frac{\int^{R_{\rm{max}}}_{0} \,T(R,u)\,\rho(R,u)\,A(R,u)\,dR}
{\int^{R_{\rm{max}}}_{0}\,\rho(R,u)\,A(R,u)\,dR} \,,
\end{equation}
where $u\equiv z/H(R)$ and $H(R)$ is the vertical scale height
given by Equation~\ref{rho_disk}.  The vertical variable, $z$, was
rescaled by $H(R)$ because lines of constant $u$ better define a
radial average due to the increase in the disk scale height with $R$
(see Equation~\ref{rho_disk}).
The function $A(R,u)$ arises because Equation~\ref{z_average_definition}
is effectively a line integral through the computational grid.  At each
$R_i$ in the discrete sum used to compute this integral, $z$ must be found
from $R_i$ and the required $u$. As different $z$ occur in the sum and
the spacing of the $z$-grid increases with $R$, we have weighted each
$z$ point by the width of the strip it effectively represents as based
on the local grid spacing; this is the origin of the $A(R,u)$ function.
Such considerations do not occur for Equation~\ref{R_average_definition}
because this integral is performed over all $z_j$ for a {\em fixed\/}
$R_i$.

The changes in the vertical temperature structure with rotation
are shown in Figures~{\ref{temp_z}} through {\ref{temp_z5}} for
spectral types B0, B3, and B5, respectively. Again, the B2 model of
Table~\ref{star_para} is similar to B3 but is not shown for brevity.
Note that for each spectral type, the set of weights $A(R,u)$ used
in Equation~\ref{z_average_definition} is always the same as the
computational grid is fixed and independent of $v_{\rm frac}$
(i.e.\ we have chosen to use the unchanging grid of Appendix~A).

In general with increasing vertical height (i.e.\ perpendicular to
the equatorial plan), the temperature is coolest in the plane of
the disk ($z=0$) and then rises to a maximum near u\,$\approx\,2$.
This maximum corresponds to the hot sheath seen in the plots of
the temperature distributions in these disks (for example, see
Figure~\ref{Temperature_panels_b0}). At larger heights, the temperature
drops below the mid-plane maximum and is nearly constant.  With increasing
rotation, several trends are seen: (1) rotation reduces the temperature
at all scale heights, (2) rotation increases the vertical extent of the cool,
inner zone, (3) rotation decreases the temperature maximum in the mid-plane, and
(4) rotation shifts the maximum temperature to larger scale heights,
although this shift is not large.

Nevertheless, there are exceptions to these general trends, particularly
in the temperatures at large $u$. In the B0 model (Figure~\ref{temp_z}),
the temperature at the upper edge of the disk decreases monotonically
with rotation. However, for the B3 model, the temperature at the upper
edge of the disk is actually largest for the greatest rotation rate.
The location of the hot sheath (the temperature maximum in $u$) moves
outward in $u$ for later spectral types; it occurs well below $u=2$ for
the B0 model at all rotation rates, whereas it is above $u=2$ for the two
later spectral types, and it moves higher with increased rotation. Hence in
the later spectral types, the temperature at the upper edge of the disk
is affected by the presence of the hot sheath.


Finally, turning back to global measures of disk temperature,
Figure~\ref{main_seq} summarizes the change in the density-weighted
average disk temperature with rotation for all four spectral types.
In general, the non-rotating models define the upper envelope of the
temperatures while the extreme rotators ($v_{\rm frac}\ge0.95$) define
the lower temperature envelope. In constructing this figure, any change
in the overall spectral type of the star due to increasing rotation has
been ignored.

\begin{figure}
\epsscale{.60}
\plotone{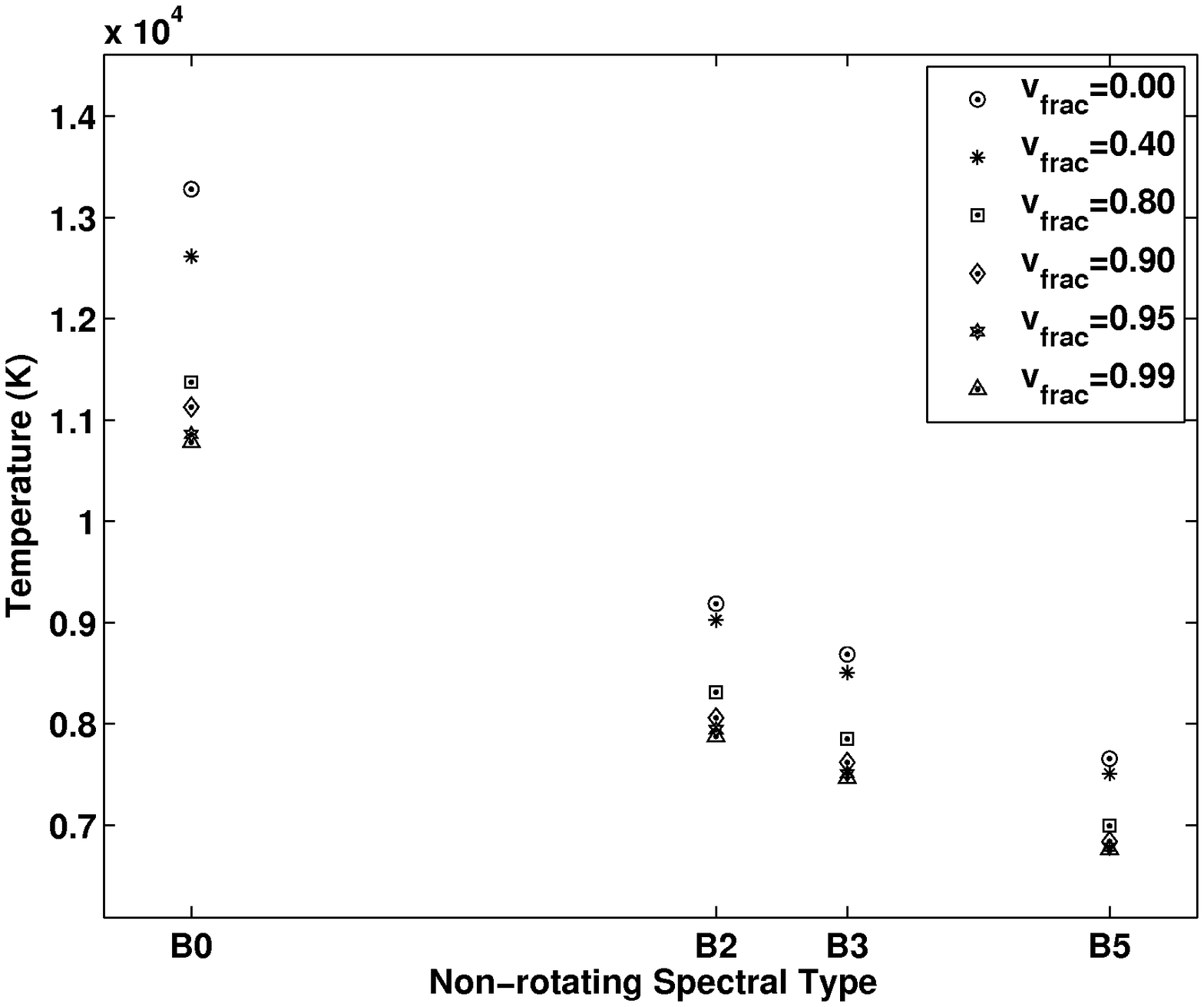}
\caption{Density-weighted temperature average plotted against spectral type
for six different rotation rates. \label{main_seq}}
\end{figure}

\section{Approximate Treatments of Gravitational Darkening}
\label{SGD/PSD}

In this section, we assess the effectiveness of partial or approximate
treatments of gravitational darkening.  In one approach, spherical
gravitational darkening (SGD), we have treated the star as a sphere and
implemented only the variation of the photospheric temperature across the
stellar surface. As this procedure is very simple to implement, it is of
interest to see how the predicted disk temperature structure compares
to the full treatment.  In another approach, pure shape distortion
(PSD), we include only the rotational distortion of the central star,
and not the temperature variation, to understand the effects of geometry
alone.  Global density and volume-weighted disk temperatures for both
approximations (SGD and PSD), and the full treatment (FGD), are shown
in Figure~\ref{discussion} for the B2 stellar model.

The differences between FGD and SGD are subtle for $v_{\rm frac}\le 0.7$,
with only small differences in the density-weighted and volume-weighted
average temperatures.  For the largest rotational rates, $v_{\rm
frac}\ge0.95$, SGD predicts density-weighted temperatures that level off
for higher rotation rates while the full treatment gives temperatures
that continue to drop.  The difference amounts to $\approx+500\,$K for
the SGD model at $v_{\rm frac}=0.99$ in the density-weighted average
temperature. In the volume-averaged temperature, the FGD temperature
actually falls below the SGD prediction in the vicinity of $v_{\rm
frac}\approx\,0.80$, but then increases above the SGD prediction for
larger rotation rates.  By $v_{\rm frac}=0.99$, the difference is again
about $500\,$K.

To further illuminate this result, these calculations were repeated with
PSD. As evident from Figure~\ref{discussion}, PSD results in essentially
no change to the average disk temperatures for $v_{\rm frac}\le0.7$
and only very small changes for faster rotation rates.  This, of course,
does not mean that the effect of the rotational distortion of the stellar
surface is unimportant; indeed the difference between the FGD treatment
and SGD noted above is the neglect of the stellar distortion. The PSD
treatment simply illustrates that the distortion of the stellar alone
produces almost no change in global disk temperatures.

However global temperature diagnostics tell only part of the story. To
illustrate how the temperature structure of the disk is reproduced
by the SGD approximation, Figure~\ref{fig:sgd_trtv} compares the
vertically-averaged (Eq.~\ref{R_average_definition}) and radially-averaged
(Eq.~\ref{z_average_definition}) disk temperatures for a B2 stellar
model computed with three rotation rates, $v_{\rm frac}=0$, $0.80$,
and $0.95$. As can be seen from the Figure, there are significant
differences between all the rotating models (either FGD or SGD) and the
non-rotating model, and all of the previously discussed effects can be
seen. Interestingly both SGD profiles for $v_{\rm frac}=0.8$ are close to
the FGD prediction. However by $v_{\rm frac}=0.95$, the SGD temperature
profiles remain close to the $v_{\rm frac}=0.8$ SGD predictions and do
not follow the trend of the full (FGD) treatment; this is particularly
noticeable in the radially-averaged profile (bottom panel) where the
(average) location of the hot sheath is very poorly predicted by the
SGD model. We conclude that SGD is an acceptable approximation to the
temperature structure only for $v_{\rm frac}\le 0.8$, and that for higher
rotation rates, the full treatment is required.

\begin{figure}
\epsscale{.60}
\plotone{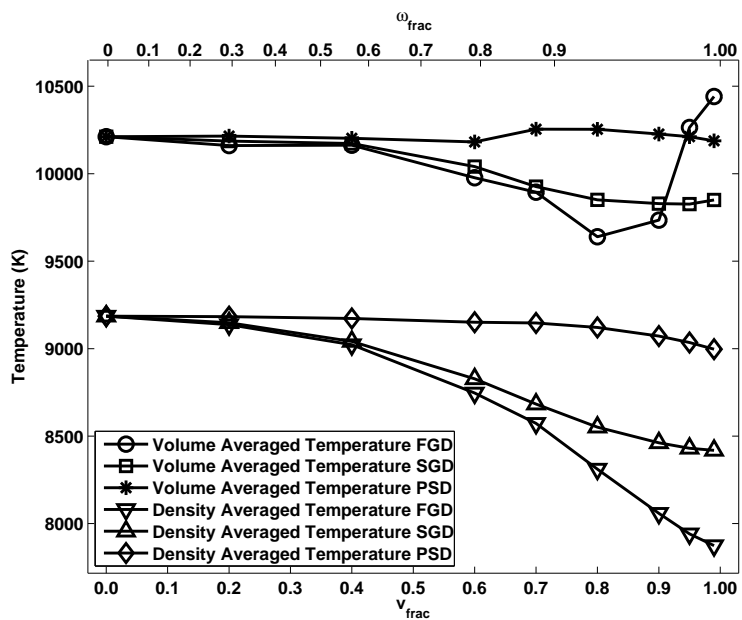}
\caption{The global volumed-averaged and density-averaged
disk temperature as a function of $v_{\rm{frac}}$ for the
B2 model. Three different treatments of gravitational darkening considered:
FGD, SGD, and PSD.
\label{discussion}} 
\end{figure}

\begin{figure}
\epsscale{.60}
\plotone{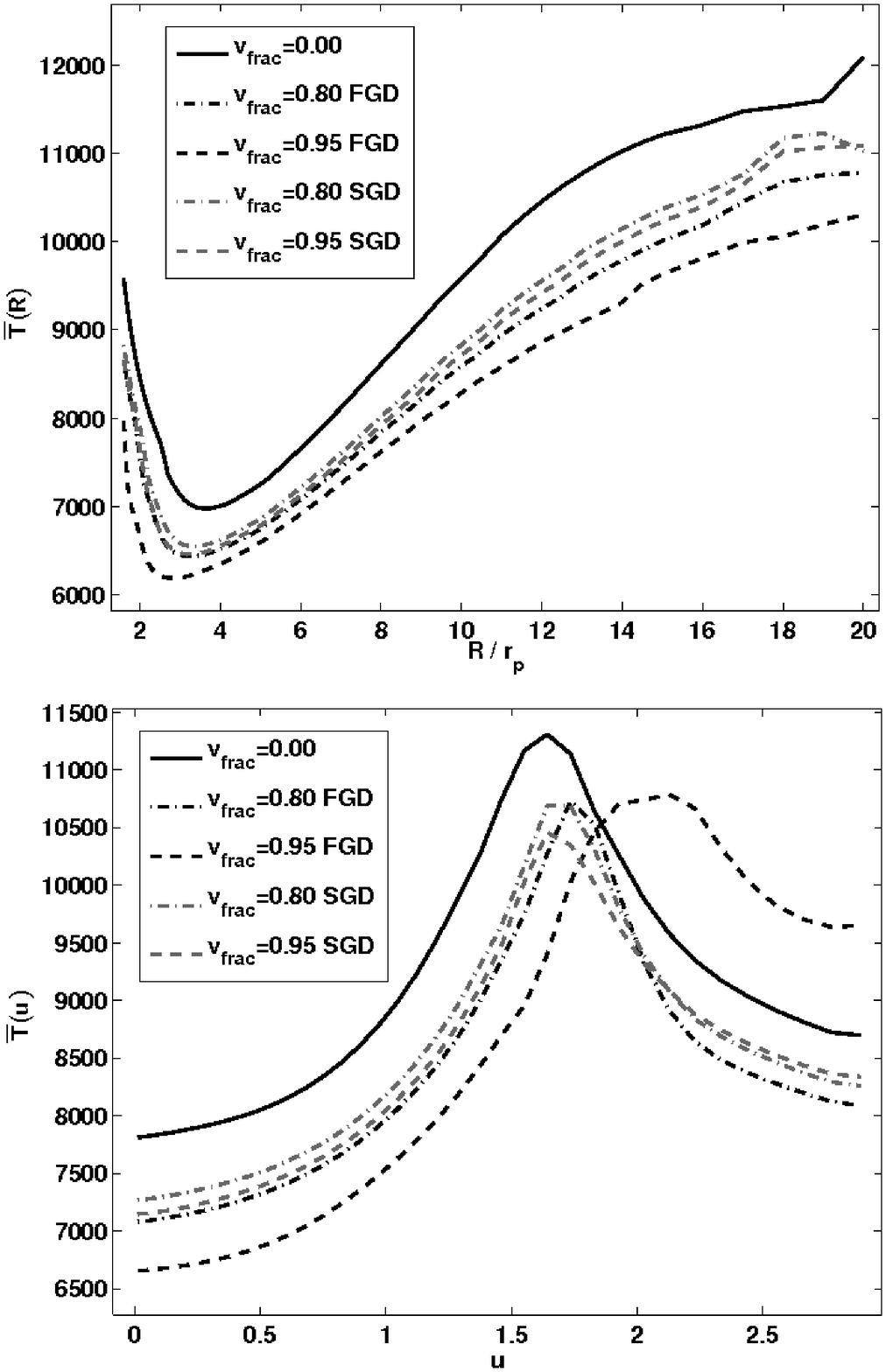}
\caption{Vertically-averaged (top panel) and radially-averaged (bottom panel)
temperature profiles (see Equations~\ref{R_average_definition} and \ref{z_average_definition})
for a B2 stellar model that compare the SGD approximation
with the full treatment (FGD). Models rotating at $v_{\rm frac}=0$ (solid lines),
$0.80$ (dot-dashed lines) and $0.95$ (dashed lines) are shown.
\label{fig:sgd_trtv}} 
\end{figure}

\section{Conclusions}
\label{conclusions}

Gravitational darkening produces noticeable changes in the temperature
structure of Be star circumstellar disks.  Rotation causes the stellar
surface temperature to diverge from a single value into a range of
temperatures, and this changes the photoionizing radiation field incident
on the disk.  Three measures of disk temperature were used to quantify the
effect of rapid rotation: detailed temperature distributions, $T(R,z)$,
vertically and radially-averaged disk temperatures, and global density
or volumed-weighted average disk temperatures that reduced the complex
temperature variations of the disk to a single, average temperature. Among
these choices, the vertically and radially-averaged temperatures are
perhaps the most useful for illustrating the effect of rapid rotation
over a wide range of rotation rates and spectral types.

Rapid rotation of the central star and the accompanying gravitational
darkening cause most of the disk gas to systematically decrease in
temperature, and this is reflected in lower density-weighted, average disk
temperatures for increasing rotation rates. However volume-averaged disk
temperatures indicate that upper regions of the disks can experience
additional heating at rotational speeds above 80\% of the critical
rate. Such changes cannot be produced by simply lowering the effective
temperature of the star with rotation.

All spectral types considered, B0--B5, show a decline in the global,
density-weighted averaged disk temperature of $\approx\,15$--$20$\%
for near critical rotation. The decline is largest for the earliest
spectral type considered, B0.  Changes in the global, volume-weighted
average disk temperature and the minimum and maximum temperatures
in the disk also occur, although results are dependent on spectral
type. The volume-weighted average temperature shows the effect of the
hot stellar pole (that develops for rapid rotation) at intermediate
spectral types B2 -- B3 (see Figures~\ref{global_temp_short_B2} and
\ref{global_temp_short_B3}).

The most important change to the temperature structure of the disks is
the expansion of the inner cool, zone close to the central star. This zone
develops in all disks of sufficient density, and this cool region generally
expands in both radius and height as the rotation rate is increased. Also
important are the hot sheaths above and below the cool, inner zone. These
generally decrease in temperature and move to larger scale heights with
increased rotation.  It is these complex changes that will most affect
physical observables such as the H$\alpha$ emission line strength or
the infrared excess. These issues will be explored in a subsequent paper.

The temperature effects of gravitational darkening on global measures of
Be star circumstellar disks can be adequately approximated by spherical
gravitational darkening (SGD, which ignores the distortion of the stellar
surface) for rotation rates less than 80\% of critical. For higher
rotation rates, the distortion of the stellar surface must be included to
accurately compute the disk's global average temperature. Comparisons of
the vertically-averaged and radially-averaged disk temperature profiles
for models with different rotation rates suggests that the general
temperature structure of the circumstellar disk can be reasonably
predicted using the SGD approximation for rotation rates of also less
than 80\% of critical.

While the effect of rapid rotation of the central star can produce complex
effects in the temperature structure of Be star circumstellar disks,
temperature is, of course, not directly accessible to observers. In a
subsequent paper, we will quantify the effect of rapid stellar rotation
on the classic diagnostics of circumstellar material, namely emission
line strengths and profiles and the predicted near-IR excess in the
system's spectral energy distribution.

\clearpage

\acknowledgements
We would like to thank the referee for many helpful comments.
This research was supported in part by NSERC, the Natural Sciences and
Engineering Research Council of Canada.  MAM acknowledges the receipt
of an Ontario Government Scholarship that funded part of this work.

\begin{appendix}
\section{Construction of the Computational Grid}
\label{app}

The primary goal of this paper is to explore changes in the thermal
structure of a circumstellar disk caused by changes in the stellar flux
due to rotation. Ideally the only difference between different models
should be the rotation speed of the star.  Rapid rotation has two effects
on the central star; the dependence of the photospheric temperature on
stellar latitude, and the Roche distortion of the stellar surface. If the
latter effect (distortion) did not occur, creating a series of comparison
models would be a simple matter of swapping out the central star and
examining the corresponding changes in the thermal structure of the disk.
Indeed, this is what occurs in the SGD approximation. Nevertheless,
the stellar surface is distorted by rotation, and this geometrical
change can be propagated into the physical properties of the disk.
The simple fact that the distance between the stellar pole and equator
changes with rotation is the central issue.  For a given point within
the disk it is possible to preserve the distance to one, but not both of
these locations. Given this, there are several different ways one could
attempt to mitigate the effects, and these approaches are discussed in
this Appendix. All of the various choices are illustrated in Figure~\ref{unbiase}
which shows in the left hand panels, a non-rotating B2 star and its associated
$(R,z)$ grid and in the right hand panels, a B2 star rotating at
$v_{\rm frac}=0.99$ and its associated $(R,z)$ grid. In each panel, the $(R,z)$
grid points in the circumstellar disk are shown as small dots.

In the most straight forward description, the grid locations $(R,z)$
within the disk are specified in terms of the equatorial radius.  In the
absence of rotation, this is same as the polar radius.  However when
$r_{\rm{eq}}$ increases due to rotational distortion, the computation grid
is stretched in both $R$ and $z$ (see Eq.~\ref{rho_disk}).  This increases
the volume and mass of the disk.  The computational grid points are also
moved further from both the stellar pole, the stellar equator, and each
other, as rotation increases.  Hence with grids constructed in this way,
any comparison between global temperature averages and total emission
for increasing rotation must be considered carefully. Because of the
simplicity of this approach, it will be referred to it as the naive grid.
It is illustrated in panels A and B of Figure~\ref{unbiase}.

A simple alternative would be to define the grid spacing and density
structure to be that of an non-rotating star, keep the grid fixed, and
then shift the whole grid and density structure outward as $r_{\rm{eq}}$
increases.  This preserves the distances of all grid points from the
stellar equator and between each other.  While the cross-sectional
area and density structure remains unchanged, both the inner and outer
radii of the disk are systematically increased causing the volume and
mass of the disk to also increase with rotation.  This effect can be
noticed when the disk emission measure is calculated.  The distance to
the pole of the star also increases systematically with rotation, and
this could bias any global temperature averages to lower temperatures.
This approach will be referred to as the physical grid and is illustrated
in panels C and D of Figure~\ref{unbiase}.

An alternative to maintaining the distance between the grid points
and the equator is to maintain the distance between grid points and the
polar axis.  Once again the grid spacing and density structure is defined
to be that of an non-rotating star, but instead of shifting it outward
as $r_{\rm eq}$ increases we simply remove sections of the grid that
would be swallowed by the star and leave the rest of the disk unchanged.
Unfortunately, the inner region which gets swallowed is also densest
region of the disk.  More of this region is removed as rotation increases,
and this systematically reduces all optical depths within the disk which
in turn increases the stellar flux reaching the outer regions. This 
approach will be referred to as the swallowed grid and is illustrated
in panels E and F of Figure~\ref{unbiase}.

One potential solution to the swallowed grid is to start the disk
at $R=(3/2)\,r_p$ which allows the central star to swell to the
inner disk boundary by critical rotation; the disk remains truly
unchanged. This preserves disk volume, mass and density structure,
and keeps the distances between each grid point and the stellar pole,
the hottest part of the star unchanged.  However, the distances to the
stellar equator still systematically change. We call this the unchanging grid
and it is illustrated in panels G and H of Figure~\ref{unbiase}.
This method, unfortunately, makes important regions of the disk empty that
were previously full and while comparisons between runs are unbiased any
comparison to previous work becomes difficult.  The dense inner region
of the disk is the source of the IR lines and a significant amount of
the mass in this region is no longer included.

In conclusion, there are four ways to define a grid surrounding an
expanding star: a naive  approach where everything increases, a physical
grid which shifts the disk, a swallowed grid in which mass is lost from
the disk; and an unchanging grid in which the star closes an inner gap.
The results included in this paper were computed with the unchanging
grid to allow the most straightforward temperature comparisons between
the difference models.

\begin{figure}
\epsscale{.80}
\plotone{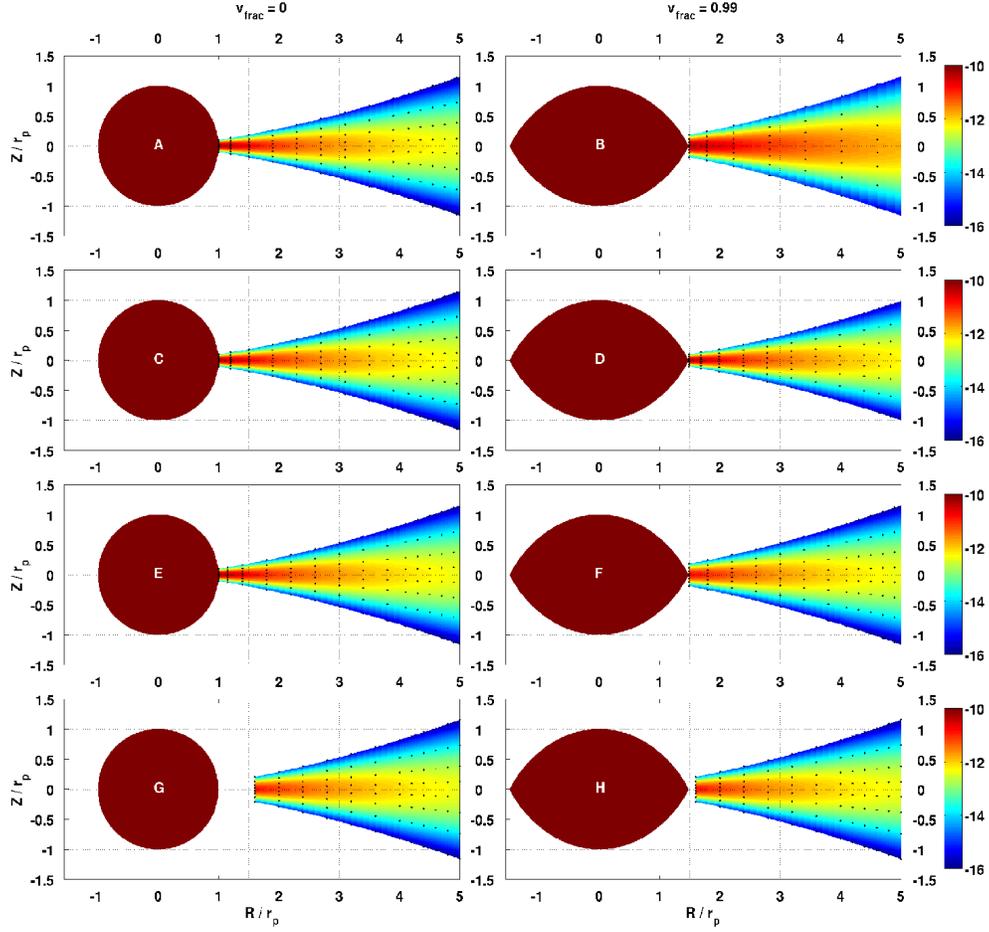}
\caption{Each panel shows a cross section of the disk with the $(R,z)$ grid
points shown as small dots.
The spherical star is not rotating, and the distorted
star is rotating at $ \omega_{\rm{frac}}=0.99$.  Panels A and B depict
the naive grid, panels C and D, the physical grid, panels E and F, the
swallowed grid, and G and H, the unchanging grid. The colours represent
the log of the disk density. \label{unbiase}} 
\end{figure}

\end{appendix}

\end{document}